\documentclass[aps,prl,twocolumn,superscriptaddress,longbibliography,10pt,floatfix]{revtex4-2}

\usepackage{amsmath,amssymb,physics,mathtools}
\usepackage{graphicx}
\usepackage{xcolor}
\usepackage{hyperref}
\usepackage{subfig}
\usepackage{placeins}

\usepackage{comment}

\newcommand{\GM}{\mathrm{GM}}
\newcommand{\q}{\mathtt{q}}

\newcommand{\AM}[1]{\textcolor{orange}{#1}}

\begin{document}

\title{Where Multipartite Entanglement Localizes:\\
The Junction Law for Genuine Multi-Entropy}

\author{Norihiro Iizuka}
\affiliation{Department of Physics, National Tsing Hua University, Hsinchu 300044, Taiwan}
\affiliation{Yukawa Institute for Theoretical Physics, Kyoto University, Kyoto 606-8502, Japan}

\author{Akihiro Miyata}
\affiliation{Yukawa Institute for Theoretical Physics, Kyoto University, Kyoto 606-8502, Japan}

\begin{abstract}
We uncover a ``junction law'' for genuine multipartite entanglement, suggesting that in gapped local systems multipartite entanglement is controlled and effectively localized near junctions where subsystem boundaries meet. Using the R\'enyi-2 genuine multi-entropy $\GM^{(\q)}_2$ as a diagnostic of genuine $\q$-partite entanglement, we establish this behavior in $(2+1)$-dimensional gapped free-fermion lattices with correlation length $\xi$. For partitions with a single junction, $\GM^{(\q)}_2$ exhibits a universal scaling crossover in $L/\xi$, growing for $L\ll\xi$ and saturating to a $\xi$-dependent constant for $L\gg\xi$, up to $\mathcal{O}(e^{-L/\xi})$ corrections. In sharp contrast, for partitions without a junction, $\GM^{(\q)}_2$ is exponentially suppressed in $L/\xi$ and drops below numerical resolution once $L\gg\xi$. We observe the same pattern for $\q=3$ (tripartite) and $\q=4$ (quadripartite) cases, and further corroborate this localization by translating the junction at fixed system size. We also provide a geometric explanation of the junction law in holography. Altogether, these results show that in this gapped free-fermion setting genuine multipartite entanglement is localized within a correlation-length neighborhood of junctions.
\end{abstract}

\maketitle

%%%%%%%%%%%%%%%%%%%%%%%%%%%%%%%%%%%%%%%%%%%%%%%%%%%%%%%%%%%%%%%%%%%%
\section{Introduction}
%%%%%%%%%%%%%%%%%%%%%%%%%%%%%%%%%%%%%%%%%%%%%%%%%%%%%%%%%%%%%%%%%%%%

Locality strongly constrains entanglement in quantum many-body systems.
For gapped states, bipartite entanglement entropy obeys the famous area law
\cite{Eisert:2008ur, Hastings:2007iok, Wolf:2007tdq, Srednicki:1993im},
reflecting the fact that entanglement is generated near entangling surfaces.
Whether an analogous organizing principle exists for
\emph{genuine multipartite entanglement} remains an open and fundamental question
\cite{Horodecki:2009zz,Guhne:2008qic}.

In this work, we answer this question affirmatively by establishing a
\emph{junction law} for genuine multi-entropy~\cite{Iizuka:2025ioc,Iizuka:2025caq}.
Genuine multi-entropy $\GM^{(\q)}$ is 
a new and sharp diagnostic of genuine $\q$-partite correlations
composed of multi-entropy \cite{Gadde:2022cqi, Penington:2022dhr, Gadde:2023zzj, Gadde:2023zni}. It isolates \emph{irreducible} $\q$-partite entanglement by subtracting all lower-partite contributions,
and vanishes for states whose correlations can be decomposed entirely into contributions involving fewer than $\mathtt{q}$ subsystems. %vanishes for states built entirely from $\tilde {\q}$-partite entanglement ($\tilde{\q}<\q$).
Using $\GM^{(\q)}$, we show that in gapped local systems genuine multipartite entanglement is effectively localized near
\emph{junctions} where subsystem boundaries meet.
We demonstrate this behavior explicitly in a solvable gapped free-fermion lattice model.
Although connected two-point correlations decay beyond the correlation length $\xi$ in gapped phases,
it is not a priori clear how \emph{irreducible} multipartite correlations organize in space.

Concretely, we study the R\'enyi-2 genuine multi-entropy $\GM^{(\q)}_2$
in $(2+1)$-dimensional free-fermion lattice models.
For tripartitions ($\q=3$) and quadripartitions ($\q=4$) containing a single junction,
we find a universal crossover controlled by the ratio $L/\xi$,
where $L$ is the system size and $\xi$ the correlation length.
For $L\lesssim \xi$, $\GM^{(\q)}_2$ grows with $L$,
while for $L\gg\xi$ it saturates to a constant determined by $\xi$ and the local opening angles at the junction.
Strikingly, when the partition geometry contains no junction,
$\GM^{(\q)}_2$ is exponentially suppressed and approaches zero in the limit $L \gg \xi$.

These results identify junctions as the spatial locus of genuine multipartite
entanglement and provide a multipartite counterpart of the entanglement area law.

Before we proceed, let us clarify why a junction is special. 
In a local state with finite correlation length $\xi$, 
connected correlations decay as 
$\langle O_X O_Y\rangle_c \sim e^{-{\rm dist}(X,Y)/\xi}$, 
and well-separated regions approximately factorize up to $O(e^{-d/\xi})$ corrections.
For a $\q$-partition with typical subsystem size $L$, 
any \emph{irreducible} $\mathtt{q}$-partite correlation must couple all $\mathtt{q}$ subsystems. 
If the boundaries do not meet, their minimal separation is $O(L)$, 
so such couplings are suppressed as $O(e^{-L/\xi})$, implying 
$\GM^{(\mathtt{q})}_n \to 0$ for $L\gg \xi$. 
A junction is therefore the unique geometry where all $\mathtt{q}$ subsystems 
approach within $O(\xi)$, allowing an $O(1)$ contribution.

%%%%%%%%%%%%%%%%%%%%%%%%%%%%%%%%%%%%%%%%%%%%%%%%%%%%%%%%%%%%%%%%%%%%
\section{Setup}
%%%%%%%%%%%%%%%%%%%%%%%%%%%%%%%%%%%%%%%%%%%%%%%%%%%%%%%%%%%%%%%%%%%%

We use the R\'enyi-$n$ genuine multi-entropy $\GM^{(\q)}_{n}$ introduced in
Refs.~\cite{Iizuka:2025ioc,Iizuka:2025caq}, which isolates irreducible $\q$-partite correlations by subtracting all lower-partite contributions. They are written in terms of multi-entropies as 
\begin{align}
&\GM^{(3)}_n(A\!:\!B\!:\!C)
= S^{(3)}_n(A\!:\!B\!:\!C) \nonumber \\
& \qquad \qquad \qquad \qquad \qquad \,\,  -\frac{1}{2}\Big(S^{(2)}_n(AB\!:\!C)+ \cdots \Big) \,,  
\label{eq:GM3_def_main} \\
&\GM^{(4)}_n(A\!:\!B\!:\!C\!:\!D)
= S_n^{(4)}(A\!:\!B\!:\!C\!:\!D) \nonumber \\
& -\frac{1}{3}\Big(S_n^{(3)}(AB\!:\!C\!:\!D) + \cdots \Big)
 +\frac{1}{3}\Big(S_n^{(2)}(ABC\!:\!D)+\cdots \Big) \nonumber \\
&\qquad \qquad \,\, \quad \qquad \qquad \qquad  \quad -a\, I_{3,n}(A\!:\!B\!:\!C\!:\!D) \,,
\end{align}
where $S^{(\q)}_n$ is the $n$-th R\'enyi multi-entropy for $\q$-partite system \cite{Gadde:2022cqi, Penington:2022dhr, Gadde:2023zzj, Gadde:2023zni}, $I_{3,n}$ is $n$-th R\'enyi triple mutual information, and $\cdots$ represents symmetrization for subsystems.
More explicit expressions for $\GM^{(\q)}_n$, $S^{(\q)}_n$ and $I_{3,n}$ are given in Appendix~A.
Throughout this work we focus on the R\'enyi index $n=2$.

%%%%%

We consider a $(2+1)$-dimensional square lattice of spinless fermions with open
boundary conditions, governed by a nearest-neighbor hopping Hamiltonian with
a staggered mass term,
\begin{equation}\label{eq:Hamiltonian}
H = -\sum_{\langle ij\rangle} c_i^\dagger c_j
+ m\sum_i (-1)^{x_i+y_i} c_i^\dagger c_i ,
\end{equation}
where $\langle i,j \rangle$ denotes a pair of nearest-neighbor sites, and $x_i, y_i \in \{1, 2, \dots, L\}$ are the $x$ and $y$ coordinates of site $i$, respectively. 
This Hamiltonian is gapped for $m\neq 0$. We assume that $m\geq 0$.
We study the many-body ground state in the half-filled sector (the Slater determinant obtained by filling all negative-energy levels, corresponding to the lowest $L^2/2$ and $(L^2-1)/2$ single-particle levels for even and odd $L$, respectively) and controlled low-lying Slater-determinant excitations obtained by applying hole creation operators. All of these states remain Gaussian and hence amenable to the correlation matrix method.

\begin{figure}[t]
\centering
\begin{minipage}[t]{0.49\columnwidth}
  \centering
  \includegraphics[width=\linewidth]{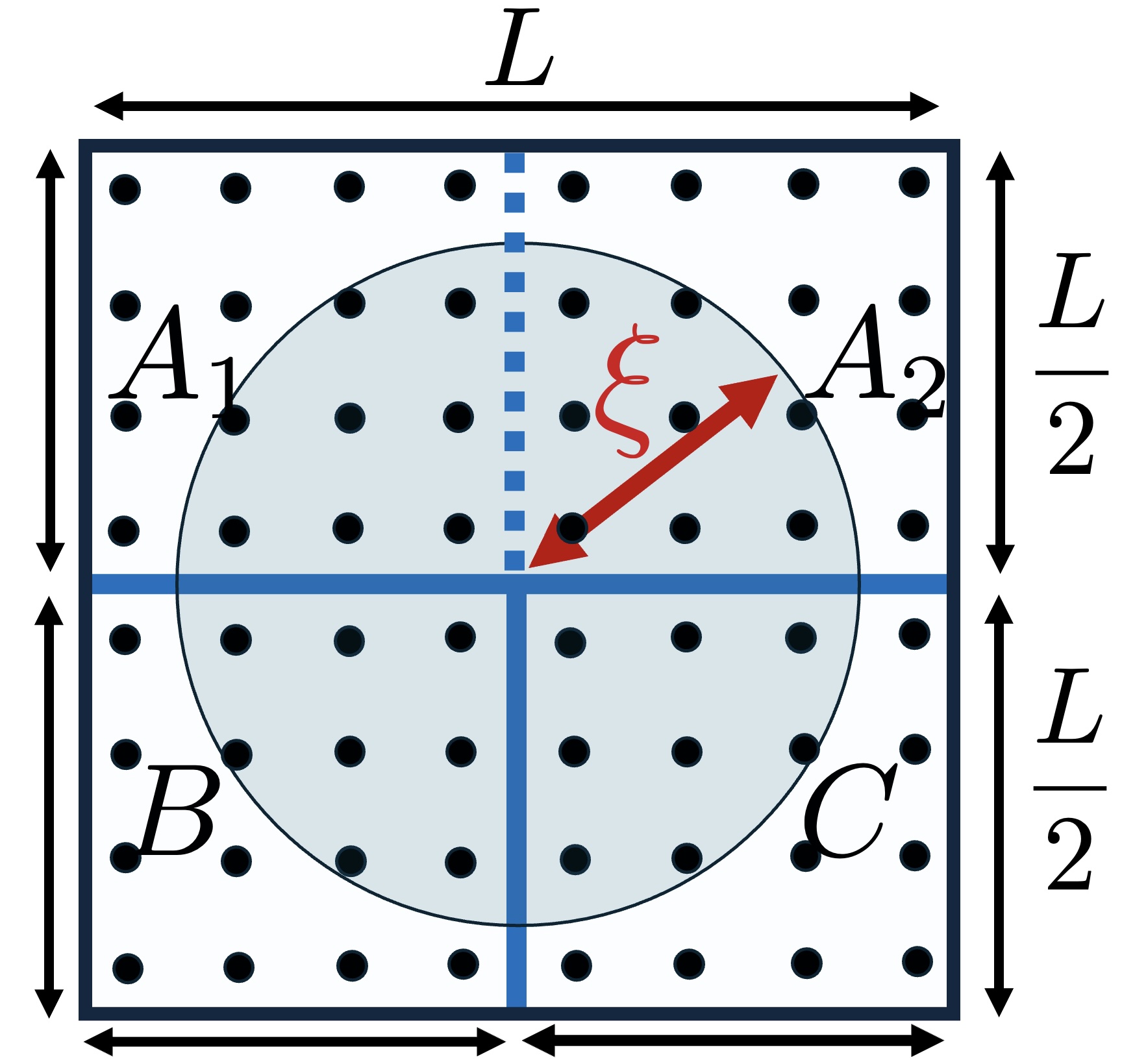}\\[-1mm]
  {\footnotesize\textbf{(a)} Junction geometry for tripartition/quadripartition.}
\end{minipage}\hspace{1mm}%
\begin{minipage}[t]{0.47\columnwidth}
  \centering
  \includegraphics[width=\linewidth]{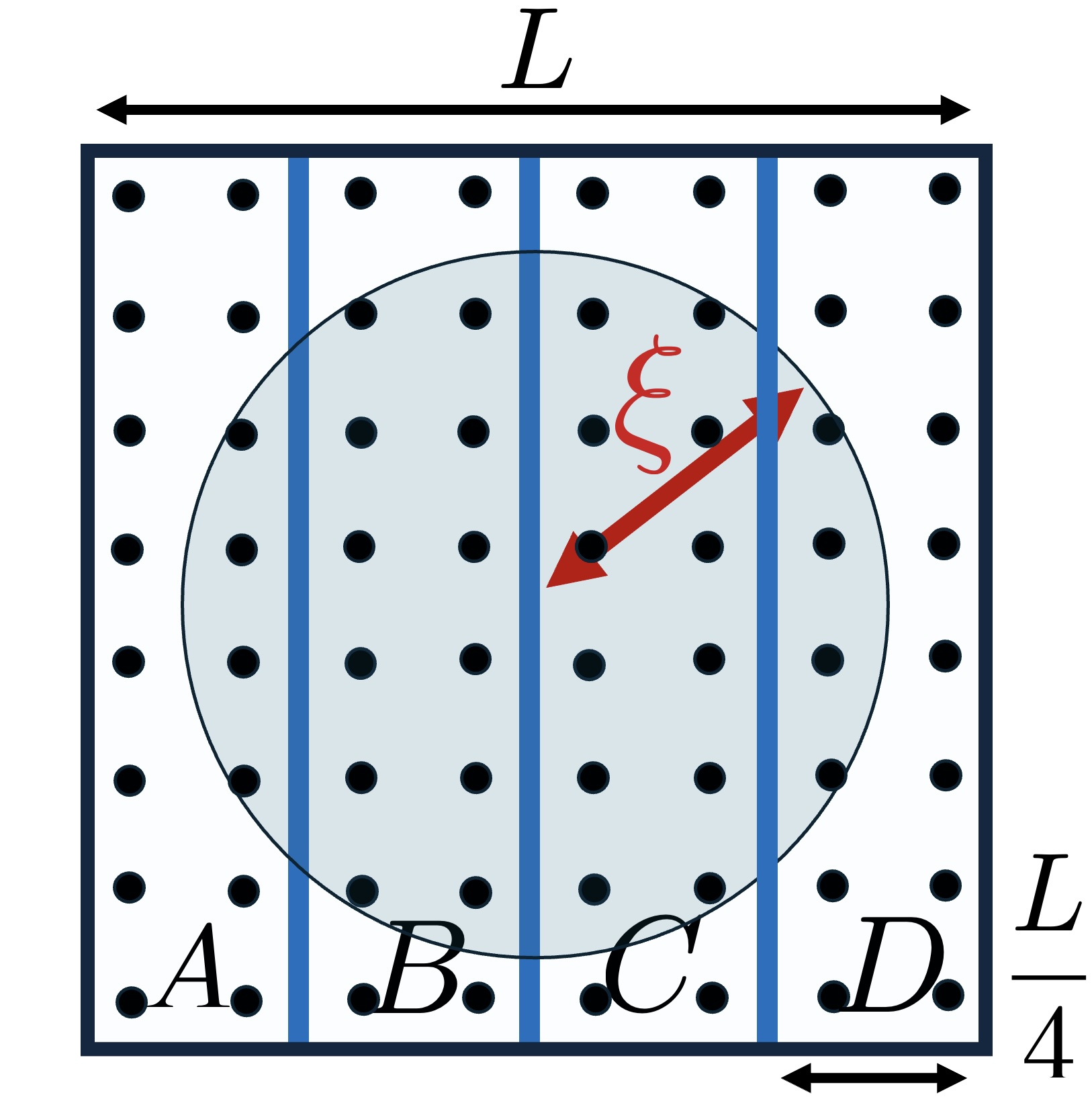}\\[-1mm]
  {\footnotesize\textbf{(b)} No-junction quadripartition.}
\end{minipage}

\caption{
\textbf{Junction vs.\ no-junction geometries.}
(a) Junction: subsystem boundaries meet at a single point; the shaded disk indicates the correlation-length scale $\xi$.
(b) No-junction: four-strip partition $A,B,C,D$ with no point adjacent to all subsystems.}
\label{fig:setup}
\end{figure}

As illustrated in Fig.~\ref{fig:setup}(a), we partition the system into subsystems whose linear sizes scale with the system size $L$
\footnote{When partition boundaries pass through lattice sites, we consistently assign the boundary degrees of freedom to the right or bottom subsystem.}.
For the tripartite case ($\q=3$), the system is divided into three subsystems
$A$, $B$, and $C$, where the region $A$ consists of two disconnected components,
$A \equiv A_1 \cup A_2$, separated by the other subsystems.
For the quadripartite case ($\q=4$), the system is divided into four subsystems
$A\equiv A_1$, $B$, $C$, and $D\equiv A_2$, each occupying one quadrant of the lattice.

In both cases, all subsystem boundaries intersect at a single point,
which we refer to as a \emph{junction} \footnote{The junction is defined as the intersection point of the subsystem boundaries in the continuum geometry and does not necessarily coincide with a lattice site. When needed, we represent it by an $\order{1}$ neighborhood of nearby sites.}.
Our central focus is on such junction geometries.
For comparison, we also study partitions in which subsystem boundaries do not
meet at a common point (no-junction geometries shown in Fig.~\ref{fig:setup}(b)).

We define the correlation length $\xi$ from the exponential decay of equal-time single-particle correlators
in the corresponding Gaussian state.
To reduce lattice anisotropy effects and the sensitivity to the precise junction location, we extract $\xi$ from a shell-averaged correlator
with the source placed near the junction,
\begin{equation}
C(r)\equiv \frac{1}{\mathcal N_r}\sum_{\substack{ \boldsymbol{j} \in\mathcal{J}\\ {\boldsymbol{i}}:\,\big||{\boldsymbol{i}}-{\boldsymbol{j}}|-r\big|\le \Delta r/2}}
\Big|\langle c_{{\boldsymbol{i}}} c^\dagger_{{\boldsymbol{j}}} \rangle\Big| 
\sim  \frac{e^{-r/\xi}}{r^{\alpha}},\label{eq:Cr_shell}
\end{equation}
where the sum averages over $\bm j$ in a small neighborhood $\mathcal J$ around the junction
and over all sites $\bm i$ at distance $r$ from $\bm j$ within a shell of thickness $\Delta r$,
with $\mathcal N_r$ the number of pairs included, and $\alpha$ accounting for the power-law correction.
We take the absolute value to suppress sign oscillations.
We then fit $\log C(r)$ over an intermediate-distance window away from the boundaries
(for details, see Appendix~B).
For the half-filled ground state, the extracted $\xi$ is consistent with the continuum estimate
$\xi\propto 1/|m|$ as $m\to 0$. For excited Slater-determinant states we use the fitted $\xi$ from \eqref{eq:Cr_shell}.

The observable of interest is the R\'enyi-2 genuine multi-entropy $\GM^{(\q)}_2$ for $\q=3,4$.
To probe a wide range of correlation lengths $\xi$, we vary the mass parameter $m$ and also consider controlled low-lying Slater-determinant excitations, which remain Gaussian, but can effectively change the correlation length $\xi$.

Although Gaussianity allows us to use the correlation-matrix method, evaluating $\q=4$ (and higher-$\q$) multi-entropies involves many multipartitions and becomes computationally heavy if implemented naively for large number of spins. 
To streamline the computation, we use a new recursion relation that expresses a $\q$-partite R\'enyi-2 multi-entropy in terms of a $(\q-1)$-partite one via canonical purification.
We summarize the recursion in Appendix~C and the resulting algorithm in Appendix~E, which generalizes the $\q=3$ techniques of Ref.~\cite{Liu:2024ulq} to arbitrary $\q$. Appendix~D provides supplementary definitions of the entanglement measures that enter the recursion,
and Appendix~F describes the explicit construction of the Gaussian states (half-filled ground state and low-lying excitations) used in our numerics.

%%%%%%%%%%%%%%%%%%%%%%%%%%%%%%%%%%%%%%%%%%%%%%%%%%%%%%%%%%%%%%%%%%%%
\section{Junction Law}
%%%%%%%%%%%%%%%%%%%%%%%%%%%%%%%%%%%%%%%%%%%%%%%%%%%%%%%%%%%%%%%%%%%%

\subsection{Partitions with a junction}
\begin{figure}[t]
\centering
\includegraphics[width=0.8\linewidth]{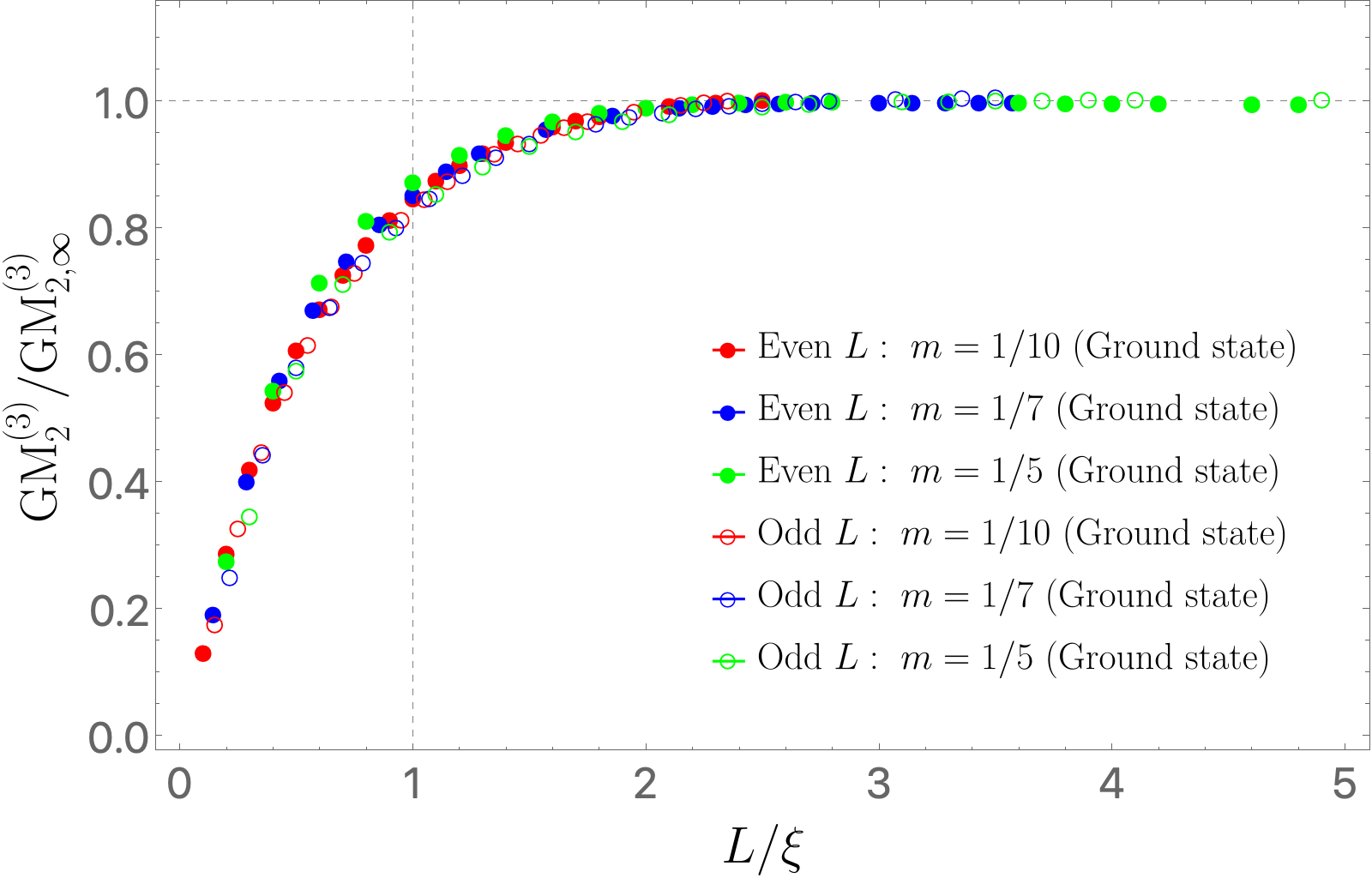}

\vspace{2mm}

\includegraphics[width=0.8\linewidth]{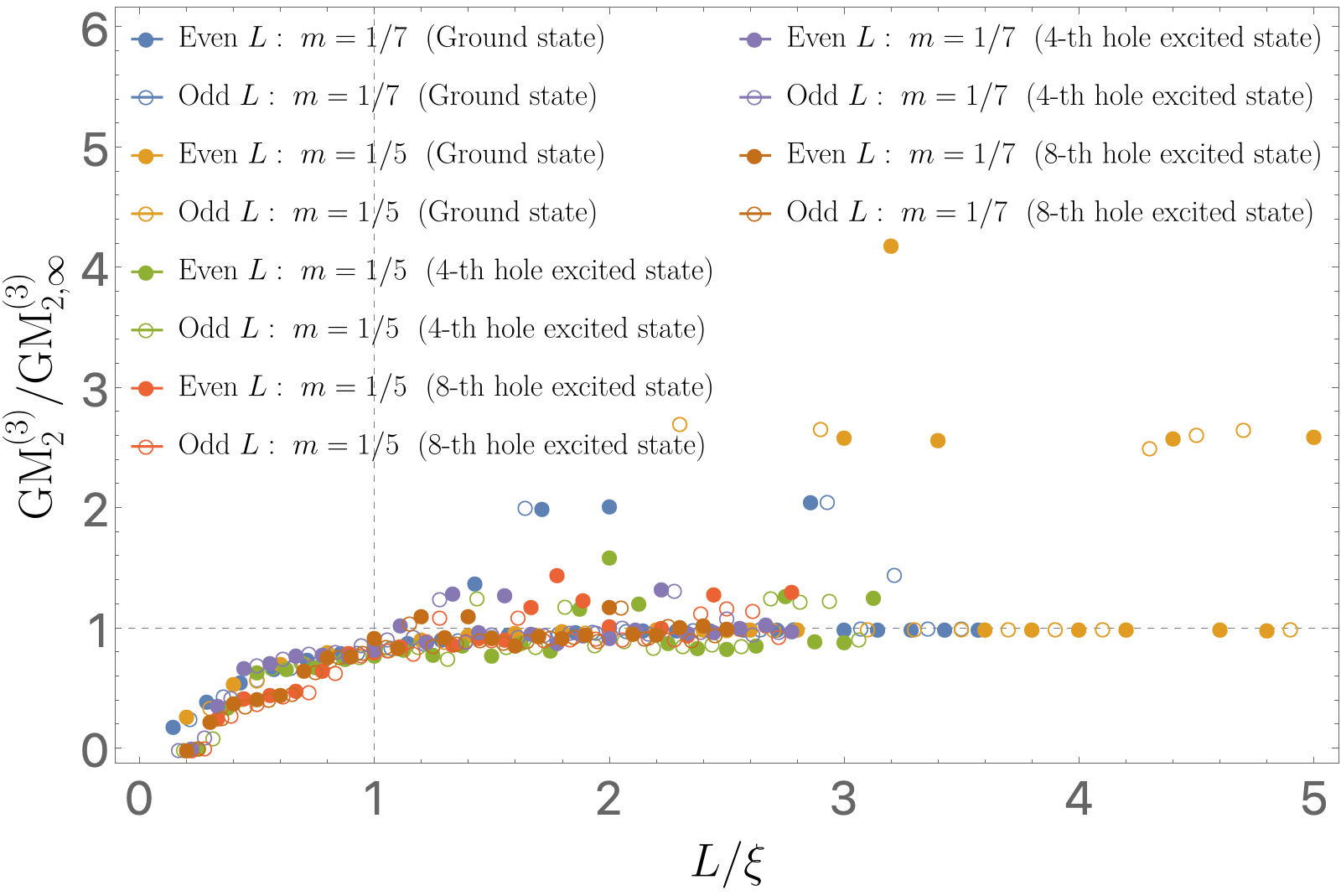}

\caption{
\textbf{Junction present: tripartite case ($\q=3$).}
Scaling collapse of the normalized genuine multi-entropy 
$\GM^{(3)}_2/\GM^{(3)}_{2,\infty}$ versus $L/\xi$
for single-junction partitions at several masses $m$.
\textbf{Top:} half-filled ground states.
\textbf{Bottom:} representative low-lying excited states.
While excited states exhibit nonuniversal deviations in the crossover regime ($L/\xi \lesssim 1$),
all data approach the same universal saturation for $L/\xi \gg 1$.
A small number of isolated points (typically $\le 5$ per series) show finite-size deviations but do not affect the scaling collapse or the large-$L/\xi$ behavior.
Full-scale plots including all data points are provided in Appendix~G.}
\label{fig:junction_q3}
\end{figure}

We first focus on the tripartite case ($\q=3$).
Figure~\ref{fig:junction_q3} summarizes our main results for partitions containing a single junction.

In the top panel of Fig.~\ref{fig:junction_q3}, we plot the genuine multi-entropy normalized by its large-$L/\xi$ saturation value,
$\GM^{(3)}_2/\GM^{(3)}_{2,\infty}$, as a function of $L/\xi$ for the half-filled ground state at several mass parameters $m$.
The data collapse onto a single curve and approach unity for $L/\xi \gg 1$,
showing a universal saturation controlled by the correlation length $\xi$.

The bottom panel of Fig.~\ref{fig:junction_q3} shows the same analysis for representative low-lying Slater-determinant excitations.
While the crossover regime ($L/\xi \lesssim 1$) displays nonuniversal deviations,
the large-$L/\xi$ saturation remains unchanged, indicating that short-distance details affect only the crossover.
We also observe a small number of outliers (a few points in each series), likely due to finite-size effects and numerical sensitivity; they do not affect the scaling collapse or the large-$L/\xi$ saturation.

\begin{figure}[t]
\centering
\includegraphics[width=0.8\linewidth]{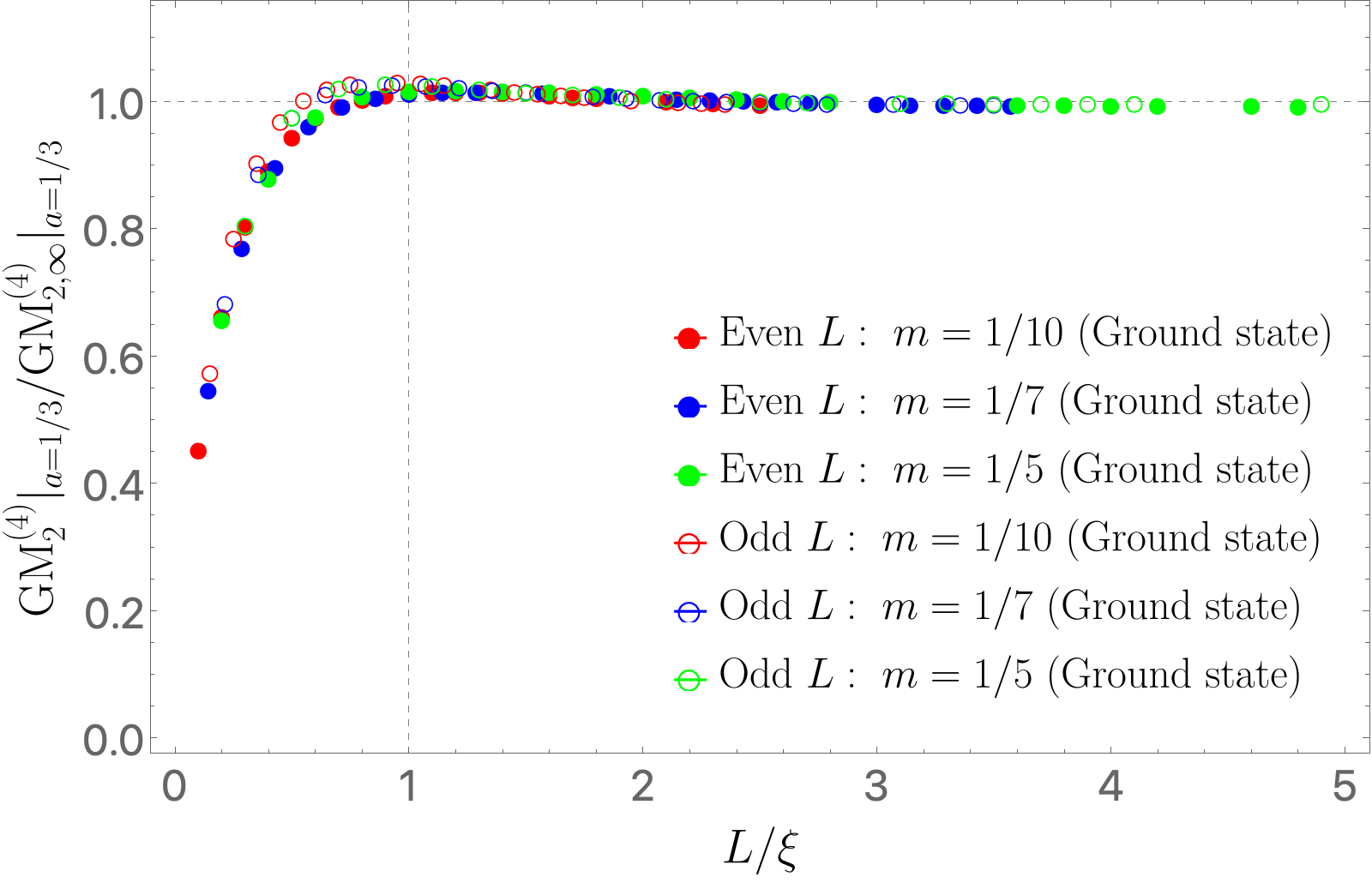}

\vspace{2mm}

\includegraphics[width=0.8\linewidth]{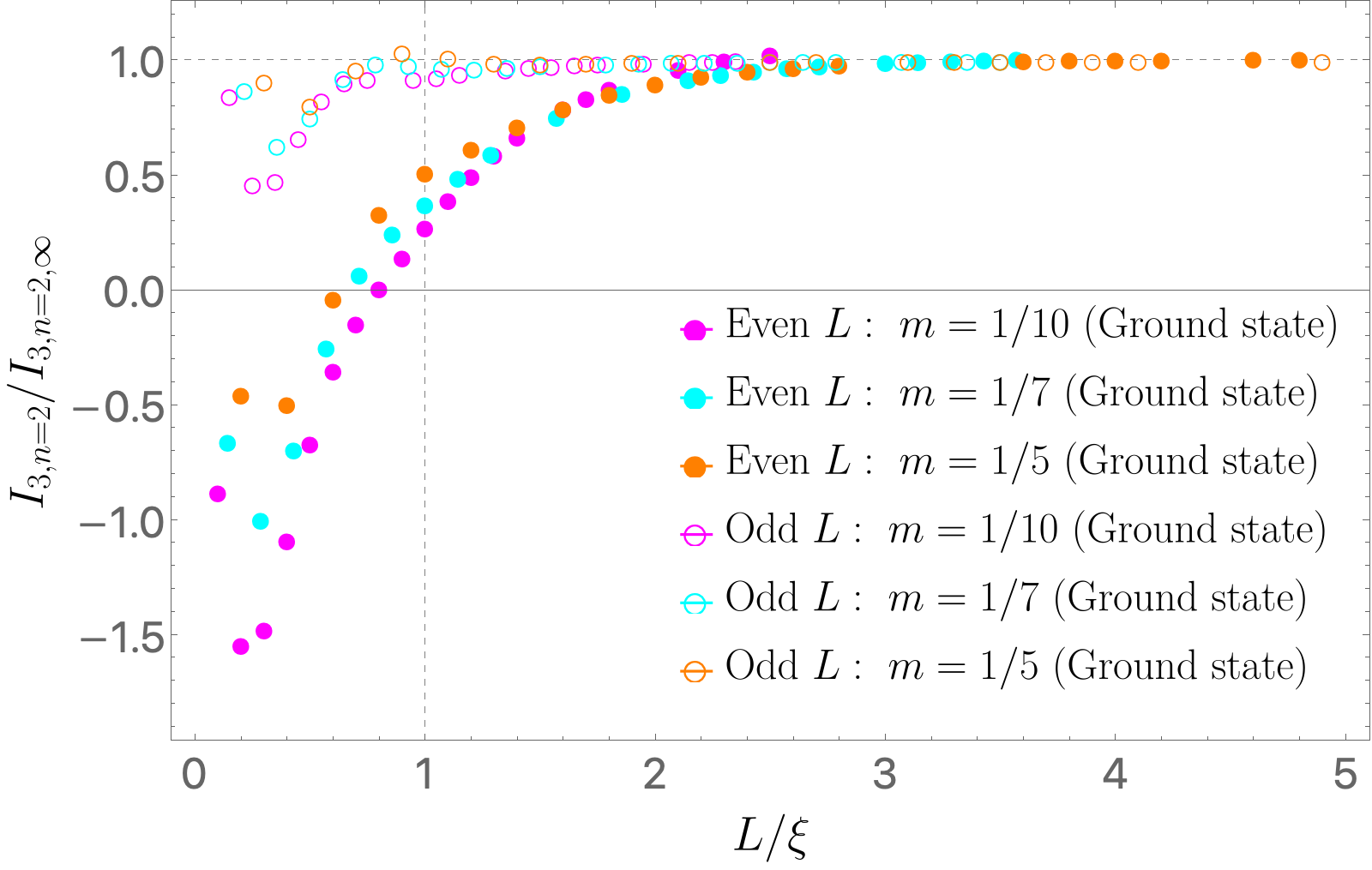}

\caption{
\textbf{Junction present: quadripartite case ($\q=4$).}
Scaling collapse of the two independent R\'enyi-2 four-partite diagnostics for single-junction quadripartitions in the ground state.
\textbf{Top:} $\GM^{(4)}_2|_{a=1/3}/\GM^{(4)}_{2,\infty}|_{a=1/3}$.
\textbf{Bottom:} $I_{3,n=2}/I_{3,n=2,\infty}$.
In both panels, data for different $m$ collapse versus $L/\xi$ and saturate at large $L/\xi$.
The bottom panel exhibits strong short-$L/\xi$ nonuniversal deviations, including negative values.
}
\label{fig:junction_q4}
\end{figure}

We now turn to the quadripartite case ($\q=4$), which provides a more stringent test of the junction law.
To reach system sizes up to $L\sim 50$, we go beyond a direct correlation-matrix implementation and use our recursion algorithm based on canonical purification (Appendix~C~and~E), which reduces the required $\q=4$ quantities to tractable $\q=3$ ones. Irreducible four-party correlations can be probed by two independent diagnostics,
$\GM^{(4)}_2|_{a=1/3}$ and $I_{3,n=2}$~\cite{Iizuka:2025caq,Balasubramanian:2014hda}.
In Fig.~\ref{fig:junction_q4} we show that both $\GM^{(4)}_2|_{a=1/3}$ and $I_{3,n=2}$ collapse as functions of $L/\xi$ across multiple masses and saturate for $L/\xi\gg1$. We note that $I_{3,n=2}$ can be negative in the crossover regime (and even at small $L/\xi$), but its sign does not affect the large-$L/\xi$ saturation. 
The agreement of these independent measures provides a stringent, basis-independent confirmation that the junction law persists beyond $\q=3$ and that genuine quadripartite entanglement is localized within an $\mathcal{O}(\xi)$ neighborhood of the junction.

\subsection{Partitions without a junction}

\begin{figure}[t]
\centering

\includegraphics[width=0.82\linewidth]{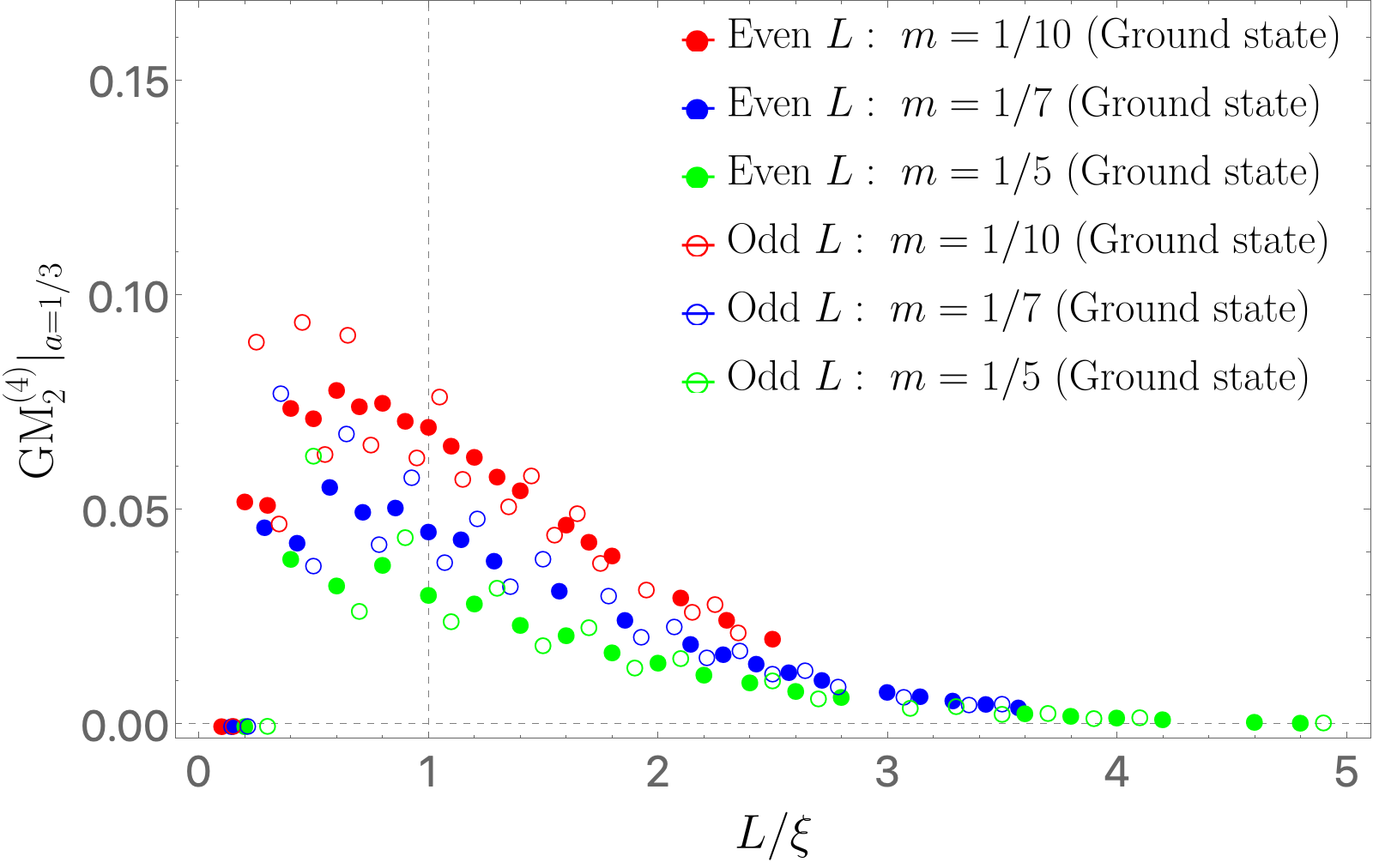}

\vspace{2mm}

\includegraphics[width=0.8\linewidth]{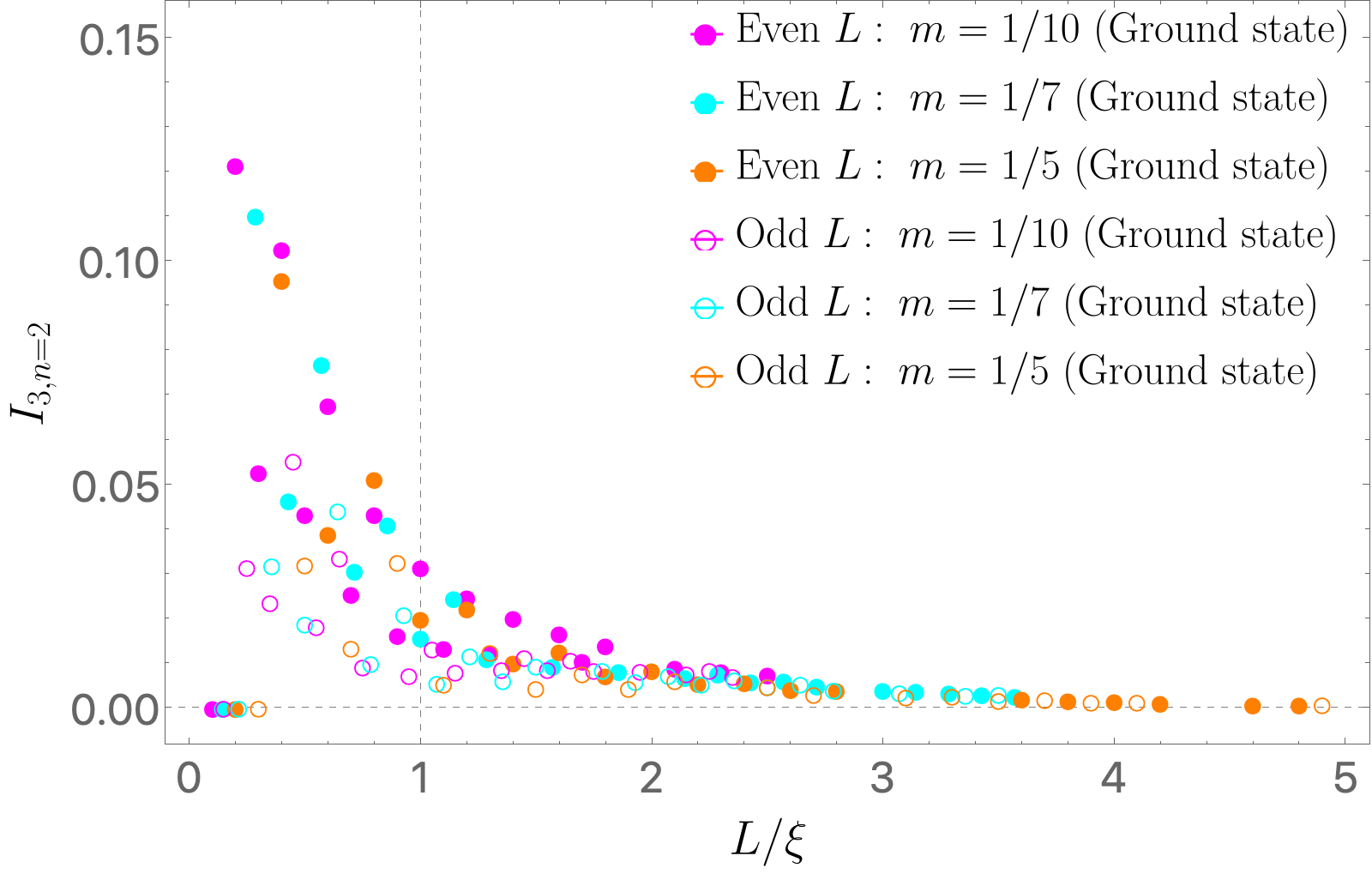}

\caption{
\textbf{No junction: genuine multipartite entanglement is exponentially suppressed.}
R\'enyi-2 quadripartite diagnostics $\GM^{(4)}_2|_{a=1/3}$ and $I_{3,n=2}$ for the junction-free partition in Fig.~\ref{fig:setup}(b).
Both decay rapidly with $L/\xi$ and are strongly suppressed for $L\gg \xi$.
}
\label{fig:nojunction}
\end{figure}

The contrast is shown in Fig.~\ref{fig:nojunction}.
For no-junction geometries, genuine four-partite correlations are exponentially suppressed in the separation scale:
both independent diagnostics, $\GM^{(4)}_2|_{a=1/3}$ and $I_{3,n=2}$, decrease rapidly with $L/\xi$ and are consistent with
\begin{equation}
\GM^{(4)}_2,\; I_{3,n=2} \;\sim\; \mathcal{O}(e^{-c\,L/\xi})
\qquad (L\gg \xi),
\end{equation}
with $c=\mathcal{O}(1)$, a geometry-dependent constant,  dropping below our numerical resolution once $L\gtrsim \xi$.
The agreement of both diagnostics confirms that the suppression is geometric, dictated by the gapped locality. Note that for $L\lesssim \xi$, the all subsystems coexist within an $\mathcal{O}(\xi)$ neighborhood, so the partition is effectively junction-like on the scale $\xi$ and the diagnostics need not be small.

\subsection{Moving junction configuration}

\begin{figure}[t]
	\centering
\includegraphics[width=0.78\columnwidth]{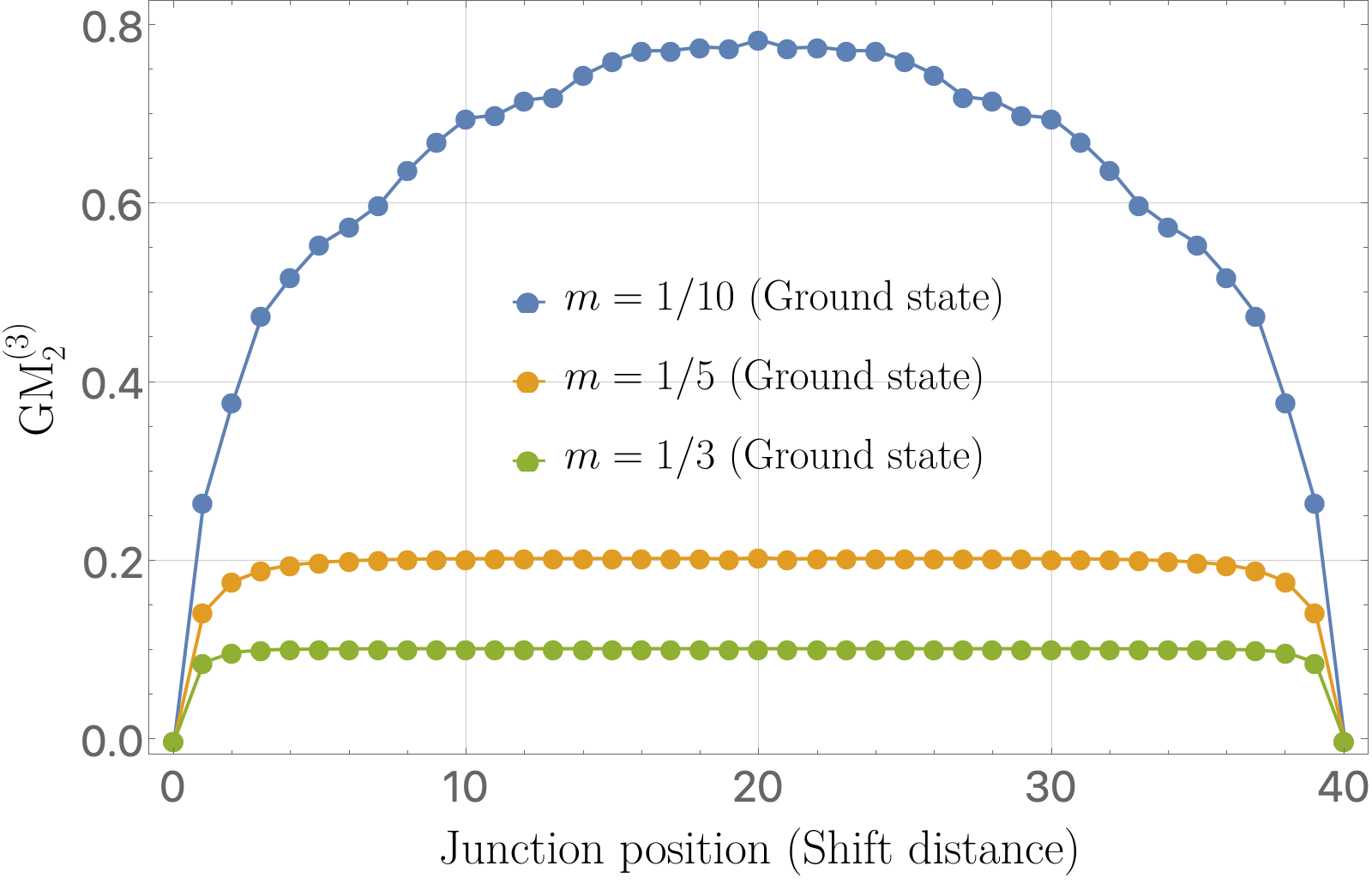}
	\caption{
\textbf{Moving junction configuration.}
$\GM^{(3)}_2$ of the half-filled ground state versus the junction position,
{\it i.e.}, the shift of the internal $B|C$ cut in Fig.~\ref{fig:setup}(a), at fixed $L=40$.
Colors label $m$.
A plateau appears when the junction is farther than $\sim \xi \,(\simeq 2/m)$ from the boundaries,
and $\GM^{(3)}_2$ is suppressed within $\sim\xi$ of a boundary.}
	\label{fig:moving_junction}
\end{figure}

As an additional spatial test of the junction law, we shift the internal $B|C$
interface (the vertical cut that extends downward from the junction in Fig.~\ref{fig:setup}(a))
while keeping $L$ fixed. Fig.~\ref{fig:moving_junction} shows that 
when the junction is farther than $\xi$ from the boundaries by more than $\xi$,
$\GM^{(3)}_2$ forms a constant plateau.
As the junction approaches the boundary within a distance $\sim \xi$,
$\GM^{(3)}_2$ is suppressed.
This confirms that irreducible multipartite correlations are localized
within an $\mathcal{O}(\xi)$ neighborhood of the junction.

%%%%%%%%%%%%%%%%%%%%%%%%%%%%%%%%%%%%%%%%%%%%%%%%%%%%%%%%%%%%%%%%%%%%%%%%
\section{Holographic picture}
%%%%%%%%%%%%%%%%%%%%%%%%%%%%%%%%%%%%%%%%%%%%%%%%%%%%%%%

\begin{figure}[t]
  \centering
  \includegraphics[width=0.48\linewidth]{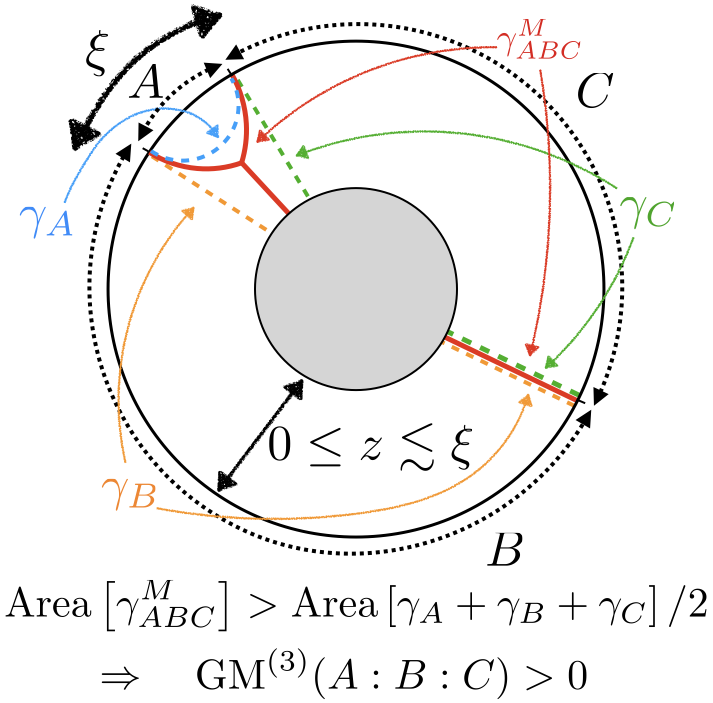}\hfill
  \includegraphics[width=0.48\linewidth]{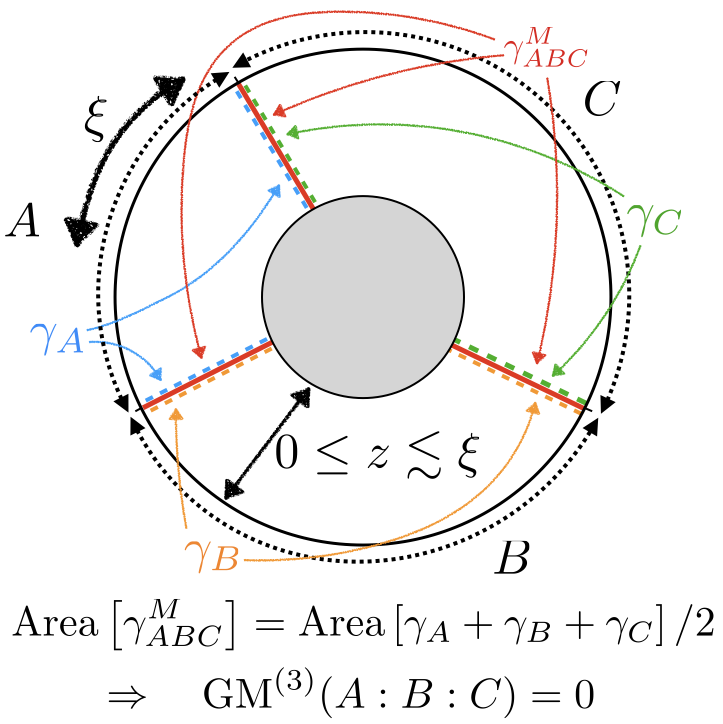}
  \caption{
    \textbf{Holographic setup with a mass gap.} The gray shaded region is excluded, corresponding to the presence of the mass gap.
    $\gamma_{ABC}^{M}$ denotes the multiway cut, and $\gamma_{\alpha}(\alpha=A,B,C)$ denotes the RT surface for a boundary subsystem $\alpha$.
    \textbf{Left:} $|A|\lesssim\xi$. This is an \emph{effective} junction configuration on the scale $\xi$, so a Y-junction forms in the bulk, resulting in $\GM > 0$.
    \textbf{Right:} $|A|\gg\xi$ case. The RT surfaces match with the multiway cut, ending on the IR cap without forming a Y-junction, resulting in $\GM \to 0$.
  }
  \label{fig:holographic_setup}
\end{figure}

Although our main results are purely non-holographic, a closely parallel statement holds in holography for genuine multi-entropy,
assuming that holographic multi-entropy is computed by the minimal multiway-cut prescription \cite{Gadde:2022cqi}.
Consider a gapped AdS/CFT setup with boundary correlation length $\xi$.
In holographic models, a mass gap is geometrically realized by a bulk spacetime that smoothly caps off in the IR, $0\leq z \lesssim \xi,$ \cite{Witten:1998zw,Klebanov:2000hb}. Accordingly, RT surfaces cannot extend beyond the IR cap,
and may end on it when the boundary region is sufficiently large \cite{Ryu:2006bv,Ryu:2006ef,Nishioka:2006gr,Klebanov:2007ws}.
As illustrated in Fig.~\ref{fig:holographic_setup}, consider a partition of the boundary \emph{without junctions}. 
If at least one boundary subsystem has linear size $\ell\lesssim\xi$, the bulk admits a nontrivial
multiway cut with a ``Mercedes--Benz'' Y-junction, and the genuine multi-entropy is positive \cite{Iizuka:2025ioc}. 
Note that for $\ell\lesssim \xi$, the partition behaves as an \emph{effective} junction configuration on the scale $\xi$, where the relevant subsystems coexist within a neighborhood $\mathcal{O}(\xi)$.
In contrast, if all subsystems satisfy $\ell \gg \xi$, the would-be Y-junction would lie behind the IR cap and cannot be realized as an extremal multiway cut,
thus the minimal multiway cut becomes junction-free and coincides with the union of RT surfaces. In this case, the genuine multi-entropy vanishes.
This provides a geometric explanation of the junction law.
%%%%%%%%%%%%%%%%%%%%%%%%%%%%%%%%%%%%%%%%%%%%%%%%%%%%%%%%%%%%%%%%%%%%%%%%%%%%%%%%%%%%%%%%%%%%%%%%%%%%%%%%
\section{Discussion}
%%%%%%%%%%%%%%%%%%%%%%%%%%%%%%%%%%%%%%%%%%%%%%%%%%%%%%%%%%%%%%%%%%%%

Our results suggest a simple organizing principle for genuine multipartite
entanglement in gapped local systems: irreducible $\q$-partite correlations
are supported only within an $O(\xi)$ neighborhood of junctions where subsystem
boundaries meet (or effectively coexist on the scale $\xi$), and are strongly suppressed otherwise.
This junction law is a multipartite counterpart of the entanglement area law,
identifying not only the scaling but the spatial locus of higher-partite correlations.

The underlying mechanism relies solely on locality and exponential clustering
in gapped phases.
Since connected correlations decay as $e^{-{\rm dist}/\xi}$,
a nonvanishing irreducible $\mathtt{q}$-partite contribution requires all $\q$ subsystems
to approach within $O(\xi)$ of a common region.
Note that this argument does not depend on Gaussianity or free-fermion structure; therefore, it suggests that the junction law extends to generic interacting
gapped systems.

Beyond gapped local phases, irreducible $\q$-partite contributions may delocalize -- e.g.\
in gapless systems (including CFTs), topologically ordered phases, or with long-range interactions.
Clarifying the minimal conditions for a junction-dominated term is an important direction.

A key next step is to test the junction law beyond Gaussian/free-fermion settings:
(i) interacting gapped lattices (tensor-network or Monte-Carlo-accessible),
(ii) bosonic systems and alternative UV regularizations, and
(iii) excited/finite-energy-density states and higher dimensions, including holographic realizations.
%%%%%%%%%%%%%%%%%%%%%%%%%%%%%%%%%%%%%%%%%%%%%%%%%

%%%%%%%%%%%%%%%%%%%%%%%%%%%%%%%%%%%%%%%%%%%%%%%%%%%%%%%%%%%%%%%%%%%%
\section*{Acknowledgments}
%%%%%%%%%%%%%%%%%%%%%%%%%%%%%%%%%%%%%%%%%%%%%%%%%%%%%%%%%%%%%%%%%%%%

We thank Shraiyance Jain for collaboration at an early stage of this project. We also thank Pochung Chen for helpful discussions.
Part of this work was presented by A.M. at the Taiwan String Workshop 2025 and the 5th ExU Annual Meeting. We thank the audience for useful questions and comments.
This work was supported in part by MEXT KAKENHI Grant-in-Aid for Transformative Research Areas A “Extreme Universe” No. 21H05184. N.I. was also supported in part by NSTC of Taiwan Grant Number 114-2112-M-007-025-MY3.

\bibliographystyle{apsrev4-2}
\bibliography{reference}

\clearpage

% ---- Appendices start here ----
\appendix

%%%%%%%%%%%%%%%%%%%%%%%%%%%%%%%%%%%%%%%%%%%%%%%%%%%%%%%%%%%%%%%%%%%%
\section*{APPENDIX}
%%%%%%%%%%%%%%%%%%%%%%%%%%%%%%%%%%%%%%%%%%%%%%%%%%%%%%%%%%%%%%%%%%%%

%%%%%%%%%%%%%%%%%%%%%%%%%%%%%%%%%%%%%%%%%%%%%%%%%%%%%%%%%%%%%%%%%%%%
%%%%%%%%%%%%%%%%%%%%%%%%%%%%

\section{Appendix A: Genuine multi-entropy}
\label{app:GM_explicit}

Explicit expressions of $\GM^{(\q)}_{n}(A_1\!:\!A_2\!:\!\dots\!:\!A_\q)$ are worked out for $\q=3$ and $\q=4$ in
Ref.~\cite{Iizuka:2025ioc}. For convenience, we collect them here:
\begin{align}
\label{eq:q3q4_genuinemulti_app}
&\GM^{(3)}_n(A\!:\!B\!:\!C)
= S^{(3)}_n(A\!:\!B\!:\!C)
-\frac{1}{2}\Big(S_n^{(2)}(AB\!:\!C) \nonumber \\ 
& \qquad \qquad \qquad \qquad \,\,+ \, S_n^{(2)}(AC\!:\!B)+S_n^{(2)}(BC\!:\!A)\Big) \\
&\GM^{(4)}_n(A\!:\!B\!:\!C\!:\!D)
= S_n^{(4)}(A\!:\!B\!:\!C\!:\!D) \nonumber \\
& \qquad -\frac{1}{3}\Big(S_n^{(3)}(AB\!:\!C\!:\!D) +S_n^{(3)}(AC\!:\!B\!:\!D) \nonumber \\
& \qquad \qquad +S_n^{(3)}(AD\!:\!B\!:\!C) +S_n^{(3)}(BC\!:\!A\!:\!D) \nonumber \\
& \qquad \qquad +S_n^{(3)}(BD\!:\!A\!:\!C)+S_n^{(3)}(CD\!:\!A\!:\!B)\Big)
\nonumber\\[-2pt]
&\qquad +\frac{1}{3}\Big(S_n^{(2)}(ABC\!:\!D)+S_n^{(2)}(ABD\!:\!C) \nonumber \\
& \qquad \qquad +S_n^{(2)}(ACD\!:\!B)+S_n^{(2)}(BCD\!:\!A)\Big) \nonumber \\
&\qquad -a\, I_{3,n}(A\!:\!B\!:\!C\!:\!D).
\end{align}
Here $S_n^{(\q)}$ denotes the $n$-th R\'enyi $\q$-partite multi-entropy, defined for the  $\q$-partite subsystems $A_1$, $A_2$, $\dots$, $A_\q$ by \cite{Gadde:2022cqi, Penington:2022dhr, Gadde:2023zzj, Gadde:2023zni}
\begin{align}
\label{themultientropydefinition1}
&\hspace{-2mm} S^{(\q)}_n(A_1:A_2:\dots:A_\q) := \frac{1}{1-n}\frac{1}{n^{\q-2}}\log \frac{Z^{(\q)}_n}{(Z^{(\q)}_1)^{n^{\q-1}}},\\
&Z^{(\q)}_n := \bra{\psi}^{\otimes n^{\q-1}} \Sigma_1(g_1)\Sigma_2(g_2)\dots\Sigma_\q(g_\q)\ket{\psi}^{\otimes n^{\q-1}},
\label{themultientropydefinition2}
\end{align}
where $\Sigma_\mathtt{k}(g_\mathtt{k})$ are twist operators for the permutation action of \(g_\mathtt{k}\) on indices of density matrices for $A_\mathtt{k}$. The action of \(g_\mathtt{k}\) can be expressed as
\begin{align}
\label{gkdefinition}
g_{\mathtt{k}} & \cdot (x_1,\dots,x_\mathtt{k},\dots,x_{\q-1}) = (x_1,\dots,x_{\mathtt{k}}+1,\dots,x_{\q-1}) \nonumber \\ &\qquad \qquad \qquad \quad \qquad \qquad \quad (\mbox{for} \,\, 1\le \mathtt{k} \le \q-1), \\
& \qquad \qquad \qquad \quad \qquad  g_\q = e ,
\end{align}
where \((x_1,x_2,\dots,x_{\q-1})\) represents an integer lattice point on a $(\q-1)$-dimensional hypercube of length \(n\) with identification of \(x_\mathtt{k}= n + 1 \) and \(x_\mathtt{k}=1\).

Note that the expression for $\GM^{(4)}_n$ includes a real parameter $a$, and its coefficient is triple mutual information $I_{3,n}$, defined as 
\begin{align}
& I_{3,n}(A\!:\!B\!:\!C\!:\!D)
= S_n^{(2)}(ABC\!:\!D)+S_n^{(2)}(ABD\!:\!C) \nonumber \\
& \qquad +S_n^{(2)}(ACD\!:\!B) +S_n^{(2)}(BCD\!:\!A)
 \\
& - S_n^{(2)}(AB\!:\!CD)-S_n^{(2)}(AC\!:\!BD)-S_n^{(2)}(AD\!:\!BC). \nonumber
\label{eq:I3n_def}
\end{align}
$I_{3,n}$ is pointed out as a diagnostic for quadripartite entanglement \cite{Balasubramanian:2014hda}. 
Thus, quadripartite genuine multi-entropy $\GM^{(4)}_n$ gives two independent diagnostics, $\GM^{(4)}_n$ with specific value of $a$ (in this paper we set $a=1/3$) and $ I_{3,n}$ \cite{Iizuka:2025caq}. The special role of $\q=3$ genuine multi-entropy was previously pointed out and studied in
~\cite{Penington:2022dhr,Harper:2024ker,Liu:2024ulq}.
An explicit formula for $\q=5$ was worked out in ~\cite{Iizuka:2025caq}.
Applications of (genuine) multi-entropy to black hole evaporation were studied in
~\cite{Iizuka:2024pzm,Iizuka:2025ioc,Iizuka:2025pqq}.

%%%%%%%%%%%%%%%%%%%%%%%%%%%%%%%%%%%%%%%%%%%%%%%%%%%%%%%%%%%%%%%%%%%%

%%%%%%%%%%%%%%%%%%%%%%%%%%%%%%%%%%%%%%%%%%%%%%%%%%%%%%%%%%%%%%%%%%%%
	\section{Appendix B: Correlation length extraction}
%%%%%%%%%%%%%%%%%%%%%%%%%%%%%%%%%%%%%%%%%%%%%%%%%%%%%%%%%%%%%%%%%%%%

	In this appendix, we provide technical details on the definition of the correlation length $\xi$.
	
	For an infinite lattice, we could define the correlation length from the asymptotic behavior of a correlation function. However, since we deal with a finite lattice system, in which the correlation functions include local information of the underlying system (e.g., boundary effects and local geometric details), it is difficult to define the correlation length in that standard way.
	
\begin{figure}[t]
	\centering
	\includegraphics[width=0.4\linewidth]{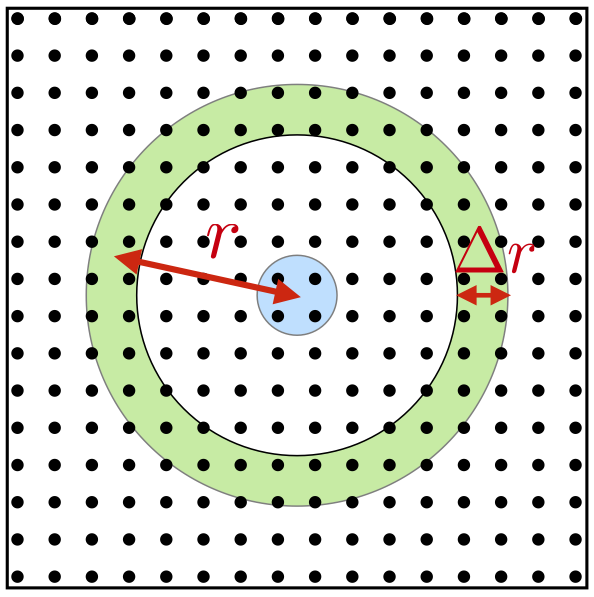}
	\caption{
		\textbf{Schematic illustration of the spatial averaging procedure defined in (\ref{eq:radial_avg}).}
		Correlations are evaluated between sites $j$ in the central blue region (around the junction) and sites $i$ located within the green shell of thickness $\Delta r$ at distance $r$. The average is taken over all sites $i$ in the green shell and all sites $j$ in the central blue region.
	}
	\label{fig:correlation_average}
\end{figure}

	Thus, rather than relying on point-to-point correlation functions, which are sensitive to local structures, we extract $\xi$ from spatially averaged correlations centered at the junction.
	Specifically, as illustrated in Figure~\ref{fig:correlation_average}, we consider the absolute value of correlation functions between sites $j$ around the junction and sites $i$ at a distance $r$ within a shell of thickness $\Delta r$, averaged over all lattice sites $i$ satisfying $\left| |i-j|-r \right|\leq \Delta r/2$:
    \begin{equation}\label{eq:radial_avg}
    C(r)\equiv \frac{1}{\mathcal N_r}\sum_{\substack{ \boldsymbol{j} \in\mathcal{J}\\ {\boldsymbol{i}}:\,\big||{\boldsymbol{i}}-{\boldsymbol{j}}|-r\big|\le \Delta r/2}}
    \Big|\langle c_{{\boldsymbol{i}}} c^\dagger_{{\boldsymbol{j}}} \rangle\Big|
    \end{equation}
	where $\mathcal N_r$ is the number of pairs included.
		This radial averaging suppresses local oscillations and isolates the typical long-distance decay behavior.
	
We fit the averaged correlation function to an exponential form with a power correction
	\begin{equation}\label{eq:correlation_fit}
		C(r) \sim \frac{e^{-r/\xi}}{r^{\alpha}}
	\end{equation}
	over an intermediate window $r_{\min} \le r \le r_{\max}$. In our preliminary numerical analysis for the masses we focus on, we observed that the exponent $\alpha$ typically takes values around $1.5$.
	Therefore, to reduce the number of free parameters and ensure a robust estimation of $\xi$, we fix $\alpha = 3/2$ in the fitting procedure.

	The lower cutoff $r_{\min}$ is chosen to avoid lattice-scale local effects, while the upper cutoff $r_{\max}$ is selected to exclude boundary contamination; since boundary effects are expected to penetrate into the bulk over a distance comparable to the correlation length $\xi$ from the edges, we restrict the fitting window to the bulk region well separated from the boundary.
	
\begin{figure}[t]
		\centering
		\includegraphics[width=0.82\linewidth]{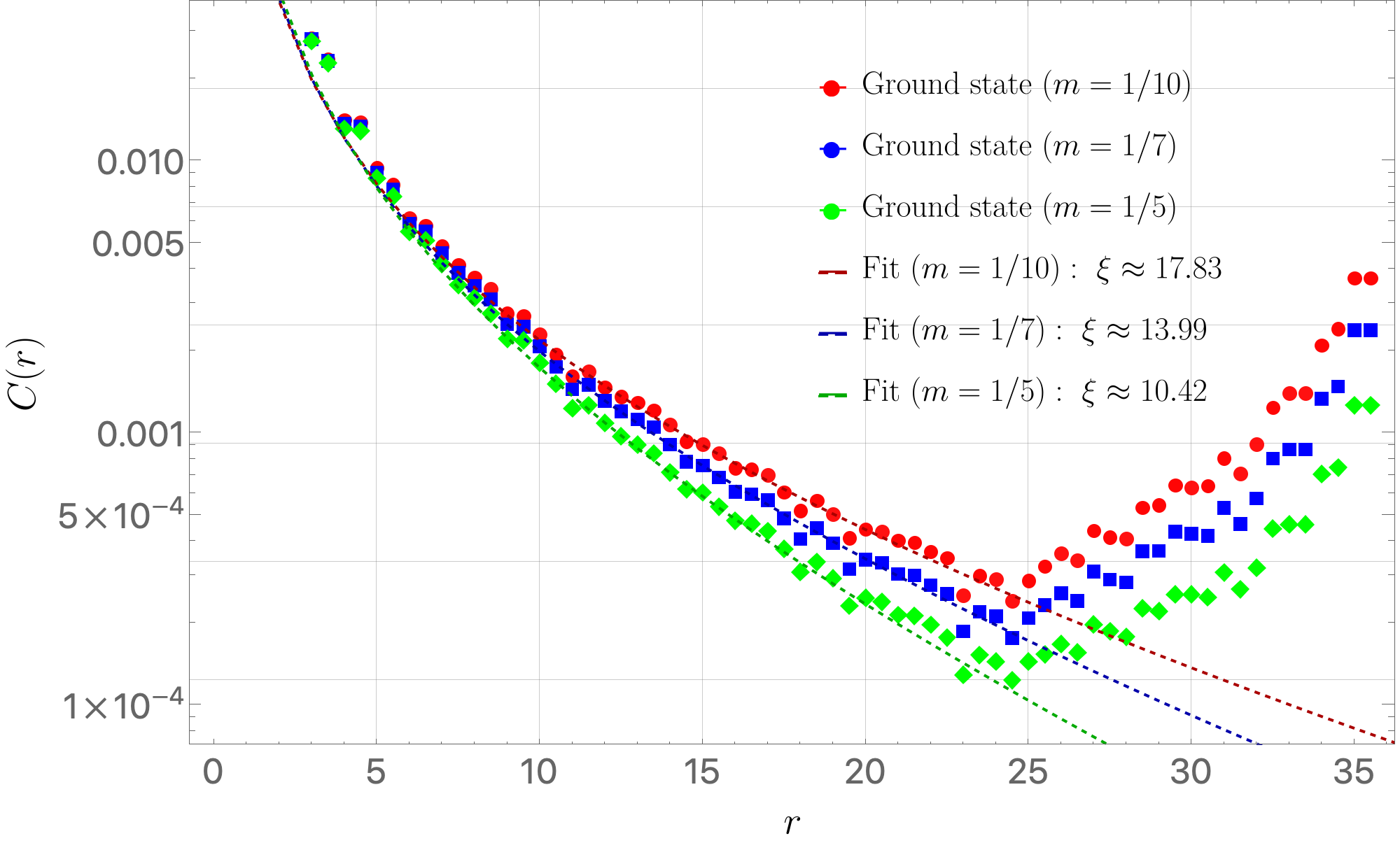}
		
		\vspace{2mm} 
		
		\includegraphics[width=0.82\linewidth]{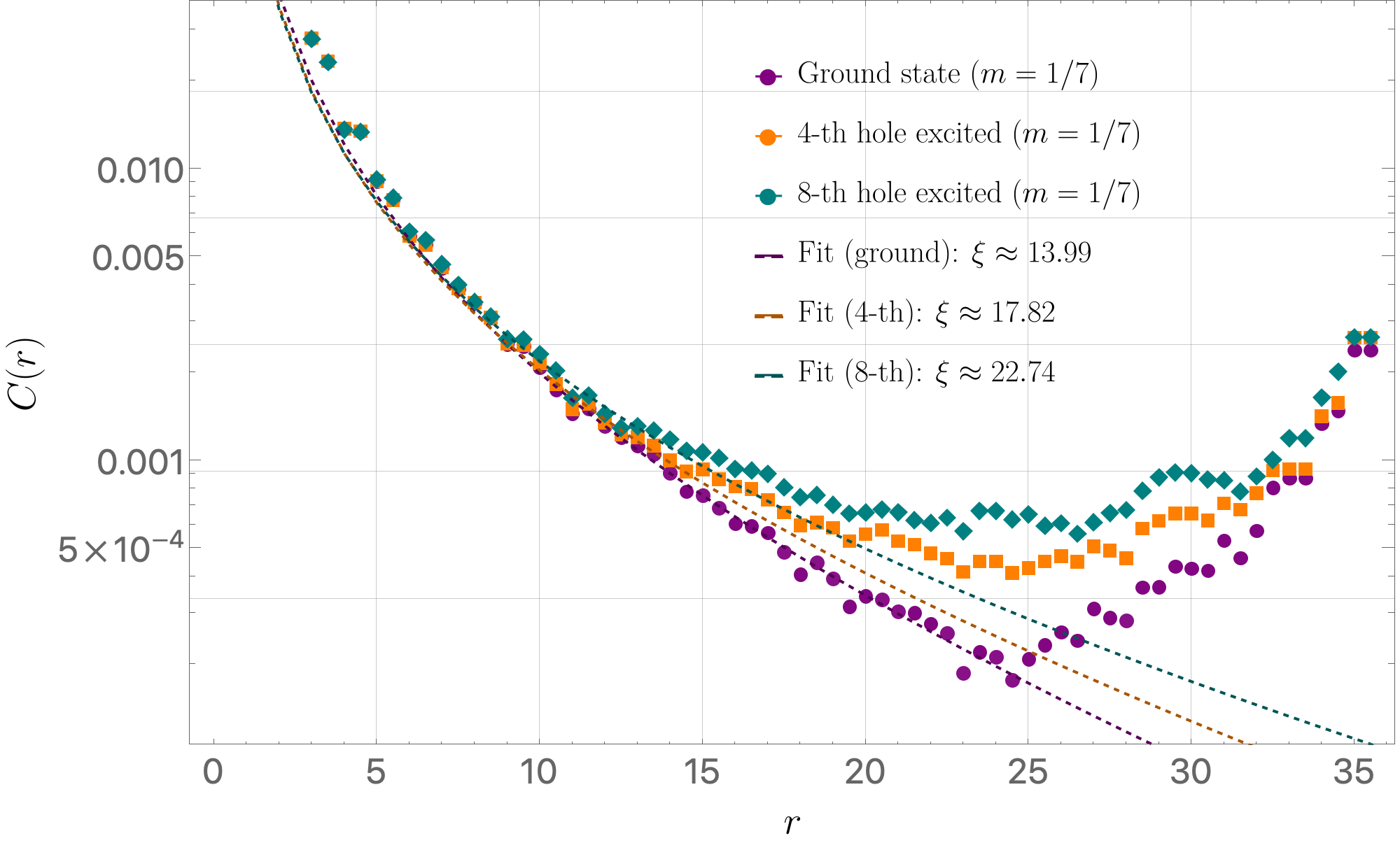}
		
		\caption{
			\textbf{Extraction of the correlation length $\xi$.}
			Spatially averaged correlation function $C(r)$ on an $L=50$ square lattice, plotted on a semi-logarithmic scale.
			The dots represent the numerical data, while the dashed lines indicate the fits to (\ref{eq:correlation_fit}) with fixed $\alpha=3/2$. We use the fitting window $r_{\min}=10$ and $r_{\max}=14.5$.
			\textbf{Top:} Ground state results for different mass parameters $m$.
			\textbf{Bottom:} Results for the ground state and hole-excited states ($m=1/7$). The correlation length $\xi$ extracted from the fit increases with the excitation level.
			In both panels, data points at large $r$ exhibit an upturn due to boundary effects (which are expected to extend over a range $\sim \xi$ from the edges) and are excluded from the fitting procedure to the extent possible.
		}
		\label{fig:correlation_fit}
	\end{figure}

	The numerical results for $C(r)$ and the corresponding fits for both the ground state and hole-excited states are shown in Figure~\ref{fig:correlation_fit}.
	The resulting $\xi$ is consistent with the crossover scale observed in the tripartite R\'enyi-2 genuine multi-entropy.
	
%%%%%%%%%%%%%%%%%%%%%%%%%%%%%%%%%%%%%%%%%%%%%%%%%%%%%%%%%%%%%%%%%%%%
\section{Appendix C: Recursion relation for $\q$-partite R\'enyi-2 multi-entropies}
\label{app:recursion}

The key identity used in our numerics is the recursion relation between $\q$-partite R\'enyi-2 multi-entropy $S_{2}^{(\q)}$ and $(\q-1)$-partite R\'enyi-2 multi-entropy $S_{2}^{(\q-1)}$
\begin{align}
&\left. S_{2}^{(\q)}\!\left(A_1: A_2: \cdots: A_{\q}\right)\right|_{\ket{\psi_{A_1\cdots A_{\q}}}} \nonumber \\
&=\frac{1}{2}\left.S_{2}^{(\q-1)}\!\left(A_1A_1^*:A_2A_2^*:\cdots: A_{\q-1}A_{\q-1}^*\right)\right|_{\ket{(\rho_{A_1\cdots A_{\q-1}})^{2/2}}} \nonumber \\
& \qquad +\left.S_{2}^{(2)}\!\left(A_1\cdots A_{\q-1}:A_{\q}\right)\right|_{\ket{\psi_{A_1\cdots A_{\q}}}}.
\label{eq:general-recursion-relation}
\end{align}
Here $\rho_{A_1\cdots A_{\q-1}}=\tr_{A_{\q}}\ket{\psi}\bra{\psi}$ is the reduced density matrix and
$\ket{(\rho_{A_1\cdots A_{\q-1}})^{2/2}}$ denotes the R\'enyi-2 canonical purification of it with auxiliary systems $A_i^*$.

Eq.~\eqref{eq:general-recursion-relation} follows from the relation shown in Ref.~\cite{Iizuka:2025elr}:
\begin{align}
&S_{R(\q-1)}^{(m=2, n=2)}\!\left(A_1: A_2: \cdots: A_{\q-1}\right) \nonumber\\
&=2\Big[S_{2}^{(\q)}\!\left(A_1: \cdots: A_{\q}\right)-S_{2}\!\left(A_1\cdots A_{\q-1}\right)\Big].
\label{eq:q-partite-MultiE-by-ReflectedE}
\end{align}
where $S_{R(\q-1)}^{(m=2, n=2)}$ is the $(m,n)$ R\'enyi reflected multi-entropy defined in Appendix D.
For $\q=3$, this reduces to the familiar relation between $(m=2,n=2)$ R\'enyi reflected entropy
and the tripartite R\'enyi-2 multi-entropy~\cite{Penington:2022dhr,Liu:2024ulq,Berthiere:2023gkx,Berthiere:2020ihq}:
\begin{equation}
\label{eq:q=3-ReflectedE-by-MultiE}
S_R^{(m=2, n=2)}\!\left(A_1: A_2\right)
=2\Big[S_2^{(3)}\!\left(A_1: A_2: A_3\right)-S_2\!\left(A_1 A_2\right)\Big].
\end{equation}

Iterating \eqref{eq:general-recursion-relation} lowers the partition number $\q$ at the expense of introducing
canonical purifications (the $*$ degrees of freedom). Repeating the procedure reduces a general $\q$-partite
R\'enyi-2 multi-entropy to a combination of tripartite and bipartite quantities on successively purified states,
which are tractable in Gaussian systems via correlation-matrix methods.
(Repeated canonical purifications have also been used as probes of multipartite structure in~\cite{Balasubramanian:2024ysu,Balasubramanian:2025jhq}.)
We note that our recursion relation enables us to express the multipartite $(m=2,n=2)$ R\'enyi reflected multi-entropy in terms of tripartite and bipartite quantities on successively purified states.

Let us set $\q=4$ in \eqref{eq:general-recursion-relation} 
\begin{equation}
    \begin{aligned}
&\left.S_{2}^{(4)}(A_1:A_2:A_3:A_4)\right|_{\ket{\psi_{A_1A_2A_3A_4}}}  \\
&=\frac{1}{2}\left.S_{2}^{(3)}(A_1A_1^{*}:A_2A_2^{*}:A_3A_3^{*})\right|_{\ket{ ( \rho_{A_1A_2A_3})^{2/2} }}\\
&\quad +\left.S_{2}^{(2)}(A_1A_2A_3:A_4)\right|_{\ket{\psi_{A_1A_2A_3A_4}}}.
\label{eq:q=4MultiE-by-q=3MultiE}
    \end{aligned}
\end{equation}
Note that bipartite (R\'enyi) multi-entropy reduces to the usual R\'enyi entropy:
\begin{equation}
\left.S_{n}^{(2)}(A_1:A_2)\right|_{\ket{\psi_{A_1A_2}}}=S_n(A_1)=S_n(A_2).
\end{equation}

Although we do not use in this paper, similarly setting $\q=5$ in \eqref{eq:general-recursion-relation}, one can obtain $\q=5$ recursion relation as 
			\begin{equation}
				\begin{aligned}
					&\left. S_{2}^{(5)}(A_1: A_2: A_3: A_4: A_5) \right|_{\ket{\psi_{A_1 A_2 A_3 A_4 A_5}}
					}\nonumber \\
					& =\frac{1}{4} \left. S_{2}^{(3)}(A_1 A_1^* A_1^{*_2} A_1^{**_2} : A_2  \cdots  : A_3 \cdots  A_3^{**_2})
					\right|_{\ket{(\rho^{(2)})^{2/2}}}  \\
					&\quad + \frac{1}{2} \left. S_{2}^{(2)}(A_1 A_2 A_3 A_1^* A_2^* A_3^* : A_4 A_4^*) \right|_{\ket{(\rho_{A_1 A_2 A_3 A_4})^{2/2}}} \\
					&\quad + \left. S_{2}^{(2)}(A_1 A_2 A_3 A_4 : A_5) \right|_{\ket{\psi_{A_1 A_2 A_3 A_4 A_5}}}.
					\label{eq:q=5MultiE-by-q=3MultiE}
				\end{aligned}
			\end{equation}
		Here, $\alpha^{*_l}$ denotes the auxiliary system introduced at the $l$-th canonical purification step, and $\rho^{(2)}$ is the reduced density matrix obtained by tracing out $A_4 A_4^*$ in $\ket{(\rho_{A_1 A_2 A_3 A_4})^{2/2}}$.

	\underline{Further Reduction: From Tripartite to Bipartite}\\
	Using the relation \eqref{eq:q=3-ReflectedE-by-MultiE}, we can further reduces tripartite R\'enyi-2 multi-entropy to the $\q=2$ quantities including the canonical purification,
		\begin{equation}\label{eq:q=3MultiE-by-ReflectedE}
			\begin{aligned}
				&\left.S_2^{(3)}\left(A_1: A_2: A_3\right)\right|_{\ket{\psi}_{A_1A_2A_3}}\\
				& =\frac{1}{2} \left. S_R^{(m=2, n=2)}\left(A_1: A_2\right) \right|_{\rho_{A_1A_2}} + \left. S_2\left(A_1A_2\right) \right|_{\rho_{A_1A_2}}\\
				&= \frac{1}{2} \left. S_{2} (A_1 A_1^*) \right|_{\ket{(\rho_{A_1 A_2})^{2/2}}} + \left. S_2\left(A_1A_2\right) \right|_{\rho_{A_1A_2}}\\
				&= \frac{1}{2} \left. S_{2}^{(2)} (A_1 A_1^* : A_2 A_2^*) \right|_{\ket{(\rho_{A_1 A_2})^{2/2}}} \\
				& \hspace{2cm} + \left. S_{2}^{(2)} (A_1 A_2 : A_3) \right|_{\ket{\psi}_{A_1 A_2 A_3}},
			\end{aligned}
		\end{equation}
		where in the last line, we used the fact that the (R\'enyi) entanglement entropy is equal to the $\q=2$ (R\'enyi) multi-entropy.

		By applying this relation \eqref{eq:q=3MultiE-by-ReflectedE} at the final stage of the recursion, $\q$-partite R\'enyi-2 multi-entropy can be fully expressed in terms of R\'enyi-2 reflected entropies and bipartite R\'enyi-2 multi-entropies involving multiple canonical purifications.  Equivalently, the quantity can be written as a combination of bipartite R\'enyi-2 multi-entropies evaluated on appropriately purified states.

		For example, through the expression \eqref{eq:q=4MultiE-by-q=3MultiE},  the quadripartite R\'enyi-2 multi-entropy can be expressed as
		\begin{equation}\label{eq:q=4MultiE-by-ReflectedE}
			\begin{aligned}
				&\left.S_{2}^{(4)}(A_1:A_2:A_3:A_4)\right|_{\ket{\psi_{A_1A_2A_3A_4}}}\\
				&=\frac{1}{4}\left.S_{R(\q=2)}^{(m=2, n=2)}(A_1A_1^{*}:A_2A_2^{*})\right|_{ \rho_{A_1A_2A_1^*A_2^*}^{(2)} }\\
				&\quad +\frac{1}{2}\left.S_{R(\q=2)}^{(m=2, n=2)}(A_1A_2:A_3)\right|_{  \rho_{A_1A_2A_3}  }\\
				&\qquad +\left.S_{2}^{(2)}(A_1A_2A_3:A_4)\right|_{\ket{\psi_{A_1A_2A_3A_4}}}\\
				&=\frac{1}{4}\left.S_{2}^{(2)}(A_1A_1^{*}A_1^{*_2}A_1^{**_{2}}:A_2A_2^{*}A_2^{*_2}A_2^{**_{2}})\right|_{\ket{ ( \rho_{A_1A_2A_1^*A_2^*}^{(2)} )^{2/2} }}\\
				&\quad +\frac{1}{2}\left.S_{2}^{(2)}(A_1A_2A_1^{*}A_2^{*}:A_3A_3^{*})\right|_{\ket{ ( \rho_{A_1A_2A_3})^{2/2} }}\\
				&\qquad +\left.S_{2}^{(2)}(A_1A_2A_3:A_4)\right|_{\ket{\psi_{A_1A_2A_3A_4}}}.
			\end{aligned}
		\end{equation}
		
		Thus, we reduce the computation of $\q$-partite R\'enyi-2 multi-entropy to that of R\'enyi-2 reflected entropies and bipartite R\'enyi-2 multi-entropies involving multiple canonical purifications.

%%%%%%%%%%%%%%%%%%%%%%%%%%%%%%%%%%%%%%%%%%%%%%%%%%%%%%%%%%%%%%%%%%%%
\section{Appendix D: Definitions of R\'enyi reflected (multi-)entropy}
\label{app:Def-Refl}

This appendix collects the minimal definitions needed for Appendix~\ref{app:recursion}C.

{\underline{Canonical purification and reflected entropy}}

Let $\rho_{AB}$ be a (generally mixed) state. %For even $m\in 2\mathbb{N}$, 
The R\'enyi canonical purification 
is the state $\ket{(\rho_{AB})^{m/2}}_{ABA^*B^*}$ ($m\in 2\mathbb{N}$) satisfying 
\begin{equation}
\label{eq:bipartite-purification-condition_short}
\tr_{A^*B^*}\!\left[\ket{(\rho_{AB})^{m/2}}\bra{(\rho_{AB})^{m/2}}\right]=(\rho_{AB})^m.
\end{equation}
where $A^*$ and $B^*$ are purifiers for $A$ and $B$. 
The $(m,n)$ R\'enyi reflected entropy \cite{Dutta:2019gen} is defined as the $n$-th R\'enyi entropy between $AA^*$ and $BB^*$ in this purified state:
\begin{align}
\label{eq:Renyi-mn-reflected-entropy_short}
&\left.S_{R}^{(m,n)}(A:B)\right|_{\rho_{AB}}
\coloneqq \left.S_{n} (AA^*)\right|_{\ket{(\rho_{AB})^{m/2}}} \\
&S_{n}(AA^*)=\frac{1}{1-n}\log\frac{\tr[\left( \rho_{AA^*}\right)^n]}{(\tr\rho_{AA^*})^n}\,, \\&\mbox{where} \quad \rho_{AA^*} \coloneqq \tr_{BB^*}\! \!\left[\ket{(\rho_{AB})^{m/2}}\bra{(\rho_{AB})^{m/2}}\right].
\end{align}
In the analytic continuation $m,n\to 1$, this reduces to the reflected entropy.

%\subsection{Reflected multi-entropy}
{\underline{Reflected multi-entropy}}

Reflected multi-entropy is a straightforward generalization of reflected entropy for higher-partite systems. The reflected entropy is defined for bi-partite system, while the reflected multi-entropy \cite{Yuan:2024yfg} is defined for a higher partite system. 

Let $\rho_{A_1\cdots A_{\q-1}}$ be a (generally mixed) state. The R\'enyi canonical purification
$\ket{(\rho_{A_1\cdots A_{\q-1}})^{m/2}}_{A_1\cdots A_{\q-1}\,;\,A_1^*\cdots A_{\q-1}^*}$ ($m\in 2\mathbb{N}$)
is defined by
\begin{align}
\label{eq:qpartite-purification-condition_short}
&\tr_{A_1^*\cdots A_{\q-1}^*}\!\left[\ket{(\rho_{A_1\cdots A_{\q-1}})^{m/2}}\bra{(\rho_{A_1\cdots A_{\q-1}})^{m/2}}\right] \nonumber \\
&=(\rho_{A_1\cdots A_{\q-1}})^m.
\end{align}
The $(m,n)$ R\'enyi reflected multi-entropy is the R\'enyi-$n$ \emph{multi-entropy} of the purified state defined by \eqref{themultientropydefinition1} and \eqref{themultientropydefinition2} in Appendix A:
\begin{align}
&\left.S_{R(\q-1)}^{(m,n)}(A_1:\cdots:A_{\q-1})\right|_{\rho_{A_1\cdots A_{\q-1}}} \nonumber \\
&\coloneqq
\left.S_{n}^{(\q-1)}(A_1A_1^*:\cdots:A_{\q-1}A_{\q-1}^*)\right|_{\ket{(\rho_{A_1\cdots A_{\q-1}})^{m/2}}}.
\label{eq:def-Renyi-mn-reflected-multi-etnropy_short}
\end{align}
In the analytic continuation $m,n\to 1$, this reduces to the reflected multi-entropy. Note that $\q=3$ case of the reflected multi-entropy is simply the reflected entropy. 

\section{Appendix E: Efficient method for $\mathtt{q}$-partite Rényi-2 multi-entropy computation}\label{app:detail-Numerics}
In this appendix, we present an \textit{efficient} method to compute an arbitrary $\q$-partite R\'enyi-2 multi-entropy, which subsequently enables the efficient evaluation of general $\q$-partite R\'enyi-2 genuine multi-entropy.
This method is a generalization of the approach in \cite{Liu:2024ulq}, in which the authors focused on tripartite genuine multi-entropy, related it to $(m=2,n=2)$ R\'enyi  reflected entropy \cite{Dutta:2019gen}, and computed the  R\'enyi  reflected entropy by using the correlation matrix method \cite{peschel2003calculation,Vidal:2002rm,Eisler:2009vye} in non-interacting fermion systems. We generalize the procedure by utilizing the recursion relation derived in Appendix~C.

As discussed in Appendix~C, an arbitrary $\q$-partite R\'enyi-2 multi-entropy can be systematically expressed in terms of the $(m=2, n=2)$ R\'enyi reflected entropy and bipartite R\'enyi-2 multi-entropy.
Since the $(m=2, n=2)$ R\'enyi reflected entropy is, by definition, the R\'enyi-$2$ entanglement entropy of a specific purified state, and the bipartite (R\'enyi) multi-entropy is equivalent to the (R\'enyi) entanglement entropy, the $\q$-partite R\'enyi-2 multi-entropy can be evaluated solely as a combination of bipartite R\'enyi-2 entanglement entropies. These quantities can then be efficiently computed through the correlation matrix method.
Without these techniques, one must generally handle matrices whose size scales exponentially with the total number of lattice sites. However, by adopting this method, we can significantly reduce the computational cost required to evaluate an arbitrary $\q$-partite R\'enyi-2 multi-entropy; the matrix sizes involved are reduced to polynomial order in the number of lattice sites.

\subsection{Correlation matrix method in a free fermion theory}

We evaluate R\'enyi-2 entanglement entropies with multiple canonical purifications using the correlation matrix method, generalizing the approach in \cite{Liu:2024ulq,Zou:2021gga} within our notation for free fermion systems.

When correlation functions satisfy Wick's theorem ({\it i.e.}, for Gaussian states), the underlying state is fully characterized by the two-point correlation matrix, allowing us to determine the reduced density matrix directly from it. 
The correlation matrix method exploits such a relation. (See e.g., \cite{Casini:2009sr,Eisler:2009vye} for general reviews on this method in the context of entanglement entropy.) By employing this approach, one can significantly reduce the computational cost required for evaluating entropies.

In the context of the present work, the half-filled ground state and its particle/hole excitations defined in Appendix~F are all Gaussian states.
Therefore, the formalism explained below is directly applicable, and the reduced density matrices are entirely determined by the correlation matrix explicitly constructed in \eqref{eq:target-correlation-matrix}.

For simplicity, we focus on the $\q=4$ case consisting of subsystems $A_1, A_2, A_3,$ and $A_4$. This case can be readily reduced to the $\q=3$ case by combining subsystems and can also be generalized to higher-partite cases. In the following, we evaluate entropies appearing in \eqref{eq:q=4MultiE-by-ReflectedE}.
  Let $c_i$ and $c_j^\dagger$ be fermionic annihilation and creation operators obeying the canonical anti-commutation relation, $\{ c_i,c_j^\dagger \}=\delta_{ij}$, and the two-point functions be
\begin{equation}
	\begin{gathered}
		\left\langle c_i c_j^\dagger  \right\rangle=C_{ij}, \quad \left\langle  c_i^\dagger c_j \right\rangle=\delta_{ij}-C_{ij},\\
		\left\langle  c_i c_j \right\rangle=\left\langle c_i^\dagger c_j^\dagger  \right\rangle=0,\\
		(i,j \in A_1,A_2,A_3,A_4)\\
	\end{gathered}
\end{equation}
where we adopted the definition of the correlation matrix $C=(C_{ij})$ in \cite{Casini:2009sr}.
We assume that all higher-point correlation functions can be computed by Wick contraction and thus expressed in terms of the above two point functions. (Note that the following discussion can be extended to cases where $\langle c_i c_j \rangle$ and $\langle c_i^\dagger c_j^\dagger \rangle$ are not necessarily zero; see, e.g., \cite{peschel2003calculation,Liu:2024ulq,Zou:2021gga}.) Let $\rho_{A_1A_2A_3A_4}$ be an underlying state (which may be either pure or mixed) to be used for computations of the correlation functions:
\begin{equation}
	C_{ij}=\tr\left[\rho_{A_1A_2A_3A_4} \, c_ic_j^\dagger\right].
\end{equation}

Since the Wick property holds, if the state $\rho_{A_1A_2A_3A_4}$ takes the form \cite{peschel2003calculation},
\begin{equation}\label{eq:4partite-density-matrix}
	\begin{gathered}
		\rho_{A_1A_2A_3A_4} =\mathcal{K}_{A_1A_2A_3A_4} \exp\left( -\mathcal{H}_{A_1A_2A_3A_4} \right)\\
		\mathcal{H}_{A_1A_2A_3A_4}=\sum_{i,j\in A_1A_2A_3A_4} h_{ij} \, c_i^\dagger c_j,
	\end{gathered}
\end{equation}
where $\mathcal{K}_{A_1A_2A_3A_4}=\tr\left[ \exp\left( -\mathcal{H}_{A_1A_2A_3A_4} \right)\right]$ is a normalization factor and $h_{ij}$ are coefficients that should be consistently determined by the correlation matrix elements $C_{ij}$. 
The explicit relation between $h=(h_{ij})$ and $C$ is give by \cite{peschel2003calculation},
\begin{equation}\label{eq:relation-C-modularH}
	\begin{aligned}
		h=-\log \left[C^{-1}-I \right] \quad \leftrightarrow  \quad C=\left( I + e^{h} \right)e^{-h}.
	\end{aligned}
\end{equation}
Let the correlation matrix $C$ be diagonalized by a unitary matrix $U$ as
\begin{equation}
	C=U\mathrm{diag}(\nu_1,\nu_2,\cdots)U^\dagger,
\end{equation}
where $\nu_\alpha$ is an eigenvalue of the matrix.
Then, the coefficient matrix $h$ is also diagonalized,
\begin{equation}
	h=U\mathrm{diag}\left(\omega_1,\omega_2,\cdots \right)U^\dagger,  \quad \omega_\alpha=-\log\left(\nu_\alpha^{-1}-1\right),
\end{equation}
and  the density matrix \eqref{eq:4partite-density-matrix} takes the diagonalized form,
\begin{equation}
	\begin{gathered}
		\begin{aligned}
			\rho_{A_1A_2A_3A_4} =\prod_\alpha \left[ \frac{1}{1+e^{-\omega_\alpha}} \exp\left( -\omega_\alpha \, d_\alpha^\dagger d_\alpha  \right)  \right],
		\end{aligned}\\
		d_\alpha= \sum_{i}U_{\alpha i}^\dagger\, c_{i}.
	\end{gathered}
\end{equation}
Using these eigenvalues, the R\'enyi entanglement entropy for the state $\rho_{A_1A_2A_3A_4}$ can be written as
\begin{equation}
	\begin{aligned}
		&S_{n}(\rho_{A_1A_2A_3A_4} )\\
		&=\frac{1}{1-n}\sum_\alpha \left[ \log\left( 1+e^{-n\omega_\alpha} \right)-n \log\left( 1+e^{-\omega_\alpha} \right)\right]\\
		&=\frac{1}{1-n}\sum_\alpha \left[ \log\left( \nu_\alpha^n +(1-\nu_\alpha)^n \right) \right]\\
		&=\frac{1}{1-n}\tr \left[ \log\left( C^n +(I-C)^n \right) \right].
	\end{aligned}
\end{equation}

While we have focused on the total system, subsystems can be treated in a similar manner. For a subsystem $A_1A_2A_3$, the associated correlation matrix $C_{A_1A_2A_3}$ is obtained by restricting $C$ to the indices $i, j \in A_1A_2A_3$,
\begin{equation}\label{eq:reduced-correlation-matrix-A1A2A3}
	C_{A_1A_2A_3}=(C_{ij})_{i,j\in A_1A_2A_3}.
\end{equation}
The reduced density matrix can be constructed from $C_{A_1A_2A_3}$,
\begin{equation}\label{eq:3partite-density-matrix}
	\begin{gathered}
		\rho_{A_1A_2A_3} =\mathcal{K}_{A_1A_2A_3} \exp\left( -\mathcal{H}_{A_1A_2A_3} \right)\\
		\mathcal{H}_{A_1A_2A_3}=\sum_{i,j\in A_1A_2A_3} \left(h_{A_1A_2A_3}\right)_{ij} \, c_i^\dagger c_j\\
		h_{A_1A_2A_3}=-\log \left[C_{A_1A_2A_3}^{-1}-I \right],
	\end{gathered}
\end{equation}
and the associated R\'enyi entanglement entropy is given by
\begin{equation}
	\begin{aligned}\label{eq:Renyi-n-EE-A1A2A3}
		&S_{n}(\rho_{A_1A_2A_3} )\\
		&=\frac{1}{1-n}\tr \left[ \log\left( (C_{A_1A_2A_3})^n +(I-C_{A_1A_2A_3})^n \right) \right].
	\end{aligned}
\end{equation}

The next step is to incorporate the R\'enyi version of the canonical purification and compute R\'enyi reflected entropy. Let us focus on  the subsystem $A_1A_2A_3$ and consider the canonical purification, which appears in \eqref{eq:q=4MultiE-by-ReflectedE}. To this end, we consider the following state,
\begin{equation}\label{eq:m-th-reduced-density-matrix}
	\begin{aligned}
		\rho_{A_1A_2A_3} ^{(m_1)}&=\frac{ \left(	\rho_{A_1A_2A_3}\right)^{m_1} }{\tr\left[  \left(	\rho_{A_1A_2A_3}\right)^{m_1} \right]}\\
		&=\frac{ \exp\left( -m_1\mathcal{H}_{A_1A_2A_3} \right)}{ \tr\left[ \exp\left( -m_1\mathcal{H}_{A_1A_2A_3} \right) \right] },\\
	\end{aligned}
\end{equation}
where $m_1\in2\mathbb{N}$. Although considering the $m_1=2$ case is sufficient for our current purposes, we present the general $m_1 \in 2\mathbb{N}$ case here for the sake of clarity.
We note that, through the relation \eqref{eq:relation-C-modularH}, the corresponding correlation matrix $C_{A_1A_2A_3}^{(m_1)}$ can be constructed by
\begin{equation}
	\begin{aligned}
		&C_{A_1A_2A_3}^{(m_1)}\\
		&=\left( I + e^{m_1 h_{(3)}} \right)e^{-m_1h_{(3)}}\\
		&=\left(C_{A_1A_2A_3}\right)^{m_1}\cdot \left[ \left(C_{A_1A_2A_3}\right)^{m_1}+ \left(I-C_{A_1A_2A_3}\right)^{m_1}   \right]^{-1}.
	\end{aligned}
\end{equation}
For the mixed state \eqref{eq:m-th-reduced-density-matrix}, we consider its canonical purification with a suitable normalization,
\begin{equation}\label{eq:canonical-purified-state-1}
	\begin{aligned}
		&\ket{ \left( \rho_{A_1A_2A_3}  \right)^{m_1/2} }_{\substack{A_1A_2A_3A_1^{*}A_2^{*}A_3^{*}}},
	\end{aligned}
\end{equation}
which satisfies the condition 
\begin{equation}
	\begin{aligned}
		\tr_{A_1^{*}A_2^{*}A_3^{*}}\left[ \ket{ \left( \rho_{A_1A_2A_3}  \right)^{m_1/2} } \bra{ \left( \rho_{A_1A_2A_3}  \right)^{m_1/2} }  \right] = \rho_{A_1A_2A_3}^{(m_1)},
	\end{aligned}
\end{equation}
and introduces three auxiliary subsystems $A_1^{*}A_2^{*}A_3^{*}$ as the (time-reversal and) particle-hole conjugate of $A_1A_2A_3$ for the purification.
This specific choice is essential for preserving the Wick property in the purified state \eqref{eq:canonical-purified-state-1}.
Here, for convenience, we consider the normalized version of the canonical purification, unlike the convention in Appendix~D.
The correlation matrix for the purified state \eqref{eq:canonical-purified-state-1} in the extended system is given by 
\begin{widetext}
	\begin{equation}\label{eq:canonical-purified-correlation-matrix-1}
		\begin{aligned}
			C_{A_1A_2A_3A_1^{*}A_2^{*}A_3^{*}}^{(m_1)} =
			\begin{pmatrix}
				C_{A_1A_2A_3}^{(m_1)} &  \sqrt{ C_{A_1A_2A_3}^{(m_1)} \left(I- C_{A_1A_2A_3}^{(m_1)} \right) } \\
				\sqrt{ C_{A_1A_2A_3}^{(m_1)} \left(I- C_{A_1A_2A_3}^{(m_1)} \right) }^{\dagger}& I-\left(C_{A_1A_2A_3}^{(m_1)} \right)^{T}
			\end{pmatrix}.
		\end{aligned}
	\end{equation}
\end{widetext}
One can verify this correlation matrix by explicitly introducing an orthogonal basis to write down \eqref{eq:m-th-reduced-density-matrix} and \eqref{eq:canonical-purified-state-1} and computing the correlation functions in the extended system while noting that $A_1^{*}A_2^{*}A_3^{*}$ is  the (time-reversal and) particle-hole conjugate of $A_1A_2A_3$.

Next, by restricting the correlation matrix \eqref{eq:canonical-purified-correlation-matrix-1} to $A_1A_2A_1^{*}A_2^{*}$
\begin{equation}\label{eq:reduced-correlation-matrix-A1A2-with-purification}
	C_{A_1A_2A_1^{*}A_2^{*}}^{(m_1)}=\left(  \left( C_{A_1A_2A_3A_1^{*}A_2^{*}A_3^{*} }^{(m_1)} \right)_{ij}  \right)_{i,j\in A_1A_2A_1^{*}A_2^{*}},
\end{equation}
the corresponding reduced density matrix is obtained as
\begin{equation}\label{eq:reduced-density-matrix-for-Purified-state}
	\begin{gathered}
		\rho_{A_1A_2A_1^{*}A_2^{*}} ^{(m_1)}=\mathcal{K}_{A_1A_2A_1^{*}A_2^{*}}^{(m_1)} \exp\left( -\mathcal{H}_{A_1A_2A_1^{*}A_2^{*}}^{(m_1)} \right)\\
		\mathcal{H}_{A_1A_2A_1^{*}A_2^{*}}^{(m_1)} =\sum_{i,j\in \substack{A_1A_2\\A_1^{*}A_2^{*}} } \left(h_{A_1A_2A_1^{*}A_2^{*}}^{(m_1)}\right)_{ij} \, c_i^\dagger c_j\\
		h_{A_1A_2A_1^{*}A_2^{*}}^{(m_1)}=-\log \left[  \left( C_{A_1A_2A_1^{*}A_2^{*}}^{(m_1)} \right) ^{-1}-I \right].
	\end{gathered}
\end{equation}
Consequently, the $(m_1,n)$ R\'enyi reflected entropy is evaluated via the Rényi entanglement entropy of this state
\begin{equation}\label{eq:Renyi-m1-n-reflected-entropy}
	\begin{aligned}
		&S_{n} \left( 	\rho_{A_1A_2A_1^{*}A_2^{*}} ^{(m_1)} \right)=S_{R}^{(m_1,n)}(A_1A_2:A_3)\\
		&=\frac{1}{1-n}\tr \left[ \log\left(  \left( C_{A_1A_2A_1^{*}A_2^{*}}^{(m_1)} \right)^n +\left(I-C_{A_1A_2A_1^{*}A_2^{*}}^{(m_1)}  \right)^n \right) \right].
	\end{aligned}
\end{equation}

By repeating this procedure, we can proceed with the remaining steps. Thus, we start with
\begin{equation}\label{eq:second-m-th-reduced-density-matrix}
	\begin{aligned}
		\rho_{A_1A_2A_1^{*}A_2^{*}} ^{(m_1;m_2)}&=\frac{ \left( \rho_{A_1A_2A_1^{*}A_2^{*}} ^{(m_1)} \right)^{m_2} }{\tr\left[  \left(	 \rho_{A_1A_2A_1^{*}A_2^{*}} ^{(m_1)} \right)^{m_2} \right]}\\
		&=\frac{ \exp\left( -m_2 \mathcal{H}_{A_1A_2A_1^{*}A_2^{*}}^{(m_1)}  \right)}{ \tr\left[ \exp\left( -m_2 \mathcal{H}_{A_1A_2A_1^{*}A_2^{*}}^{(m_1)}  \right) \right] },\\
	\end{aligned}
\end{equation}
and the corresponding correlation matrix
\begin{equation}
	\begin{aligned}
		&C_{A_1A_2A_1^{*}A_2^{*}}^{(m_1;m_2)}\\
		&=\left( 	C_{A_1A_2A_1^{*}A_2^{*}}^{(m_1)} \right)^{m_2}\\
		& \qquad \cdot   \left[ \left( 	C_{A_1A_2A_1^{*}A_2^{*}}^{(m_1)}  \right)^{m_2}+ \left(I- 	C_{A_1A_2A_1^{*}A_2^{*}}^{(m_1)} \right)^{m_2}   \right]^{-1}.
	\end{aligned}
\end{equation}
Then, as in \eqref{eq:canonical-purified-state-1}, we again canonically purify the state \eqref{eq:second-m-th-reduced-density-matrix} by introducing four auxiliary subsystems $A_1^{*_2}A_2^{*_2}A_1^{**_2}A_2^{**_2}$ as the (time-reversal and) particle-hole conjugate of $A_1A_2A_1^{*}A_2^{*}$. Similar to \eqref{eq:second-m-th-reduced-density-matrix}, we obtain the corresponding correlation matrix
\begin{widetext}
	\begin{equation}\label{eq:canonical-purified-correlation-matrix-2}
		\begin{aligned}
			C_{A_1A_2A_1^{*}A_2^{*}A_1^{*_2}A_2^{*_2}A_1^{**_2}A_2^{**_2}}^{(m_1;m_2)} =
			\begin{pmatrix}
				C_{A_1A_2A_1^{*}A_2^{*}}^{(m_1;m_2)} &  \sqrt{ C_{A_1A_2A_1^{*}A_2^{*}}^{(m_1;m_2)} \left(I- C_{A_1A_2A_1^{*}A_2^{*}}^{(m_1;m_2)} \right) } \\
				\sqrt{ C_{A_1A_2A_1^{*}A_2^{*}}^{(m_1;m_2)} \left(I-  C_{A_1A_2A_1^{*}A_2^{*}}^{(m_1;m_2)} \right) }^{\dagger}& I-\left( C_{A_1A_2A_1^{*}A_2^{*}}^{(m_1;m_2)} \right)^{T}
			\end{pmatrix}.
		\end{aligned}
	\end{equation}
\end{widetext}
Thus, again by restricting the correlation matrix to $A_1A_1^{*}A_1^{*_2}A_1^{**_2}$ and constructing the reduced density matrix through the relation \eqref{eq:relation-C-modularH}, we obtain the R\'eny entanglement entropy
\begin{widetext}
	 \begin{equation} \label{eq:Renyi-m2-n-reflected-entropy}
	 	\begin{aligned}
	 		&S_{n} \left( 	\rho_{A_1A_1^{*}A_1^{*_2}A_1^{**_2}} ^{(m_1;m_2)} \right)=\left.S_{R}^{(m_2,n)}(A_1A_1^{*}:A_2A_2^{*})\right|_{ 	\rho_{A_1A_2A_1^{*}A_2^{*}} ^{(m_1)} }\\
	 		&=\frac{1}{1-n}\tr \left[ \log\left(  \left( C_{A_1A_1^{*}A_1^{*_2}A_1^{**_2}}^{(m_1;m_2)} \right)^n +\left(I-C_{A_1A_1^{*}A_1^{*_2}A_1^{**_2}}^{(m_1;m_2)}  \right)^n \right) \right].
	 	\end{aligned}
	 \end{equation}
\end{widetext}

Finally, for a quadripartite pure state $\rho_{A_1A_2A_3A_4}= \ket{\psi}\bra{\psi}$, by combining the relation \eqref{eq:q=4MultiE-by-ReflectedE} and the explicit expressions for the correlation matrices \eqref{eq:Renyi-n-EE-A1A2A3}, \eqref{eq:Renyi-m1-n-reflected-entropy} and \eqref{eq:Renyi-m2-n-reflected-entropy}, the quadripartite R\'enyi-2 multi-entropy is expressed in terms of correlation matrices,
\begin{widetext}
	\begin{equation}\label{eq:q=4-Renyi-2-multiE-Correlation-matrix}
		\begin{aligned}
			S_{2}^{(4)}(A_1:A_2:A_3:A_4)&= -\frac{1}{4}\tr \left[ \log\left(  \left( C_{A_1A_1^{*}A_1^{*_2}A_1^{**_2}}^{(2;2)} \right)^2 +\left(I-C_{A_1A_1^{*}A_1^{*_2}A_1^{**_2}}^{(2;2)}  \right)^2 \right) \right]\\
			& \qquad  -\frac{1}{2}\tr \left[ \log\left(  \left( C_{A_1A_2A_1^{*}A_2^{*}}^{(2)} \right)^2 +\left(I-C_{A_1A_2A_1^{*}A_2^{*}}^{(2)}  \right)^2 \right) \right]\\
			&\qquad \qquad  -\tr \left[ \log\left( (C_{A_1A_2A_3})^2 +(I-C_{A_1A_2A_3})^2 \right) \right].
		\end{aligned}
	\end{equation}
\end{widetext}

While we have focused on the $\q=4$ case thus far, the expression for $\q=3$ can be readily obtained using \eqref{eq:q=3MultiE-by-ReflectedE},  \eqref{eq:Renyi-n-EE-A1A2A3} and \eqref{eq:Renyi-m1-n-reflected-entropy}. More explicitly, for a tripartite pure state $\rho_{\tilde{A}_1\tilde{A}_2\tilde{A}_3} = \ket{\psi}\bra{\psi}$,  tripartite R\'enyi-2 multi-entropy is expressed as
 \begin{equation}\label{eq:q=3-Renyi-2-multiE-Correlation-matrix}
	\begin{aligned}
		&S_{2}^{(3)}(\tilde{A}_1:\tilde{A}_2:\tilde{A}_3)\\
		&=  -\frac{1}{2}\tr \left[ \log\left(  \left( C_{\tilde{A}_1\tilde{A}_1^{*}}^{(2)} \right)^2 +\left(I-C_{\tilde{A}_1\tilde{A}_1^{*}}^{(2)}  \right)^2 \right) \right]\\
		&\qquad \qquad  -\tr \left[ \log\left( (C_{\tilde{A}_1\tilde{A}_2})^2 +(I-C_{\tilde{A}_1\tilde{A}_2})^2 \right) \right].
	\end{aligned}
\end{equation}
For comparison, we note that, for a bipartite pure state $\rho_{\hat{A}_1\hat{A}_2} = \ket{\psi}\bra{\psi}$,  bipartite R\'enyi-2 multi-entropy (which is equal to R\'enyi-2 multi-entropy) is given by
 \begin{equation}\label{eq:q=2-Renyi-2-multiE-Correlation-matrix}
	\begin{aligned}
		&S_{2}^{(2)}(\hat{A}_1:\hat{A}_2)=S_{2}(\hat{A}_1)\\
		&=-\tr \left[ \log\left( (C_{\hat{A}_1})^2 +(I-C_{\hat{A}_1})^2 \right) \right].
	\end{aligned}
\end{equation}

By utilizing these formulae, one can efficiently evaluate R\'enyi-2 multi-entropies and, consequently, Rényi-2 genuine multi-entropies for large systems.

%%%%%%%%%%%%%%%%%%%%%%%%%%%%%%%%%%%%%%%%%%%%%%%%%%%%%%%%%%%%%%%%%%%%

%%%%%%%%%%%%%%%%%%%%%%%%%%%%%%%%%%%%%%%%%%%%%%%%%%%%%%%%%%%%%%%%%%%%
\section{Appendix F: Gaussian state construction}\label{app:construction-states}
%%%%%%%%%%%%%%%%%%%%%%%%%%%%%%%%%%%%%%%%%%%%%%%%%%%%%%%%%%%%%%%%%%%%

To access a wider range of effective correlation lengths, we consider
low-lying excited Gaussian states obtained by hole (or particle) excitations
around the half-filled ground state.

To explicitly construct these states, we first prepare the many-body ground state, known as the half-filled ground state, in the half-filled sector by the Slater determinant obtained by filling all negative-energy levels, corresponding to the lowest $L^2/2$ and $(L^2-1)/2$ single-particle levels for even and odd $L$, respectively.
Concretely, after diagonalizing the hopping Hamiltonian with the mass term \eqref{eq:Hamiltonian} by a unitary matrix $W$, we have
\begin{equation}
	\begin{gathered}
		H=\sum_{\alpha} \varepsilon_{\alpha} d^{\dagger}_{\alpha}d_{\alpha}, \quad 
		c_{i}=\sum_{\alpha} W_{i \alpha }d_{\alpha},
	\end{gathered}
\end{equation}
where $\varepsilon_{\alpha}$ are the eigenvalues of the single-particle Hamiltonian $\mathcal{H}$ defined via the coefficient matrix of $H=\sum_{i,j}\, \mathcal{H}_{ij} c^{\dagger}_{i} c_{j}$ from \eqref{eq:Hamiltonian}: $\mathcal{H}=W\operatorname{diag}\left\{ \varepsilon_{\alpha} \right\}W^{\dagger}$. 
For notational simplicity, we order the $L^2$ eigenvalues $\varepsilon_{\alpha}$ and index them by integers $\alpha$ such that negative indices correspond to occupied states and non-negative indices to unoccupied states:
\begin{gather}
	\varepsilon_{\alpha} \leq  \varepsilon_{\beta} \quad \text{for} \quad \alpha < \beta, \\
	\alpha \in \{\cdots, -2, -1\} \quad \text{if} \quad \varepsilon_{\alpha} < 0, \nonumber \\
	\alpha \in \{0, 1, 2, \cdots\} \quad \text{if} \quad \varepsilon_{\alpha} \geq 0.
\end{gather}
(Note that in the gapless case $m=0$, we include the zero-energy modes in the non-negative index sector).
Then, the half-filled ground state is given by
\begin{equation}\label{eq:hf-ground-state}
	\ket{\Psi_{0}}= \prod_{\alpha <0 }\, d^{\dagger}_{\alpha} \ket{0},
\end{equation}
where $\ket{0}$ is the vacuum state annihilated by the annihilation operators.

Then, we define excited (Slater-determinant) states by occupying the lowest
positive-energy modes or removing particles from the highest occupied
negative-energy modes.
	Specifically, we define the $l$-th \textit{hole} excited state by 
	\begin{equation}
		\begin{aligned}
				\ket{\Psi_{-l}} &\coloneqq d_{-l}\cdots d_{-2}\, d_{-1} \ket{\Psi_{0}} \\
				&= \prod_{\alpha <-l }\, d^{\dagger}_{\alpha} \ket{0}
		\end{aligned}\qquad l=1,2,\cdots
	\end{equation}
	and the $m$-th \textit{particle} excited state by
	\begin{equation}
		\begin{aligned}
			\ket{\Psi_{m}} & \coloneqq d^{\dagger}_{m-1}\cdots d^{\dagger}_{1}\, d^{\dagger}_{0} \ket{\Psi_{0}} \\
			& = \prod_{\alpha <m }\, d^{\dagger}_{\alpha} \ket{0}
		\end{aligned}\qquad m=1,2,\cdots.
	\end{equation}
		One can organize these states in the form
	\begin{equation}\label{eq:n-th-state}
			\,\, \ket{\Psi_{n}}=\prod_{\alpha <n }\, d^{\dagger}_{\alpha} \ket{0}\qquad n=\cdots, -2,-1,0,1,2,\cdots.
	\end{equation}
	
	All of these states remain Gaussian and are therefore accessible via the
	correlation matrix method explained in Appendix E.
	In particular, the input correlation matrix can be written as
	\begin{equation}
		\begin{aligned}
			\left(C_{(n)}\right)_{ij}&\coloneqq \bra{\Psi_n} c_i \,  c^\dagger_j\ket{\Psi_n}\\
			&= \sum_{\alpha \geq n} W_{i\alpha} W^{\dagger}_{\alpha j}\\
			&=\sum_{\alpha \geq n} \braket{i}{\varepsilon_{\alpha} }\braket{\varepsilon_{\alpha} }{j},
		\end{aligned}
	\end{equation}
	where, in the last line, we introduced the eigenvector $\ket{\varepsilon_{\alpha}}$ for the single-particle Hamiltonian $\mathcal{H}$ with the eigenvalue $\varepsilon_{\alpha}$, and the site basis $\ket{i}$.
	Thus, we can prepare the correlation matrix $C_{(n)}$ by the projection operator 
	\begin{equation}\label{eq:target-correlation-matrix}
		C_{(n)}=\sum_{\alpha \geq n} \ket{\varepsilon_{\alpha} }\bra{\varepsilon_{\alpha} }.
	\end{equation}

\begin{figure}[t]
		\centering
		\includegraphics[width=0.7\columnwidth]{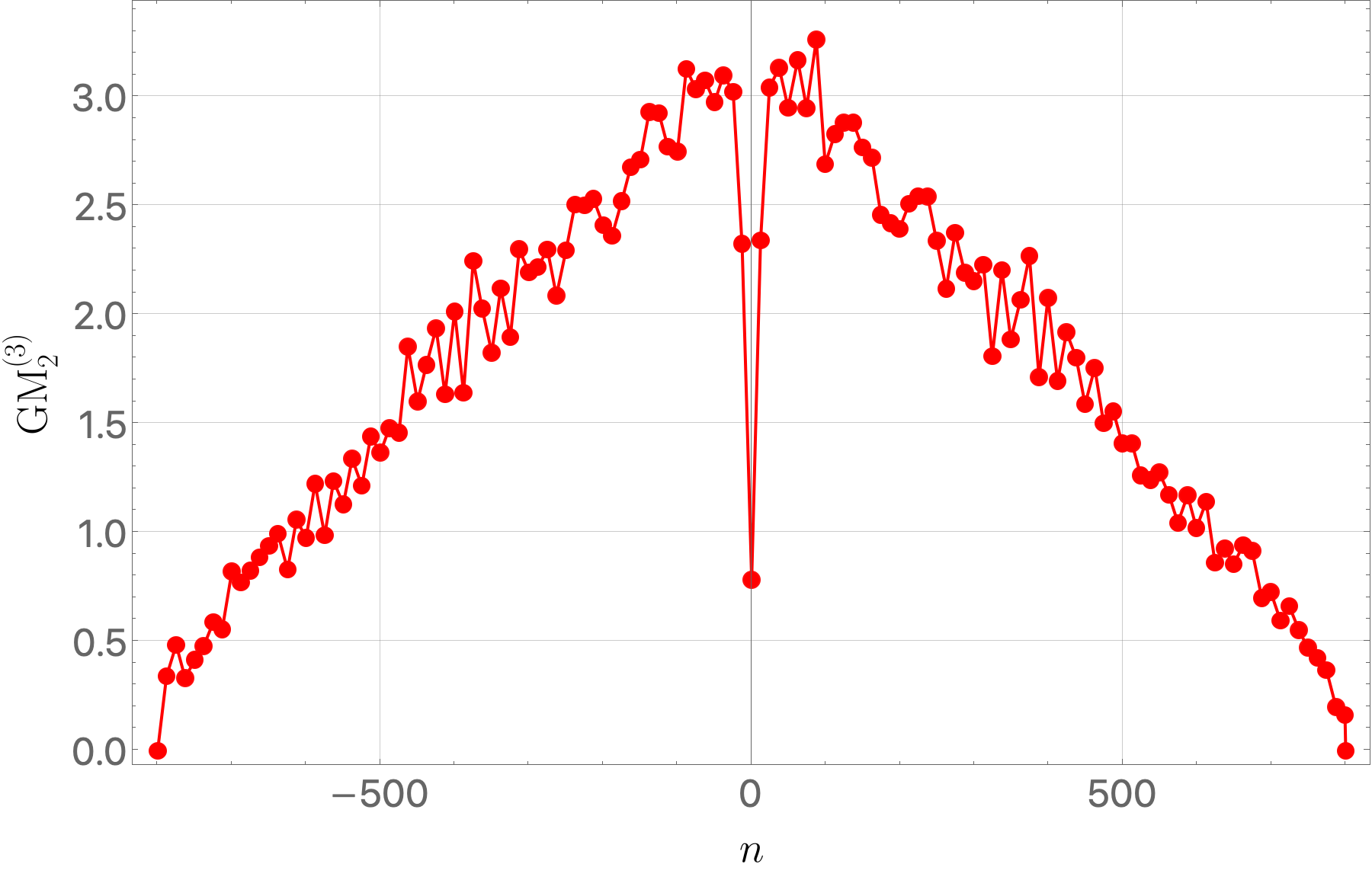} \\
		\vspace{2mm}
		\includegraphics[width=0.7\columnwidth]{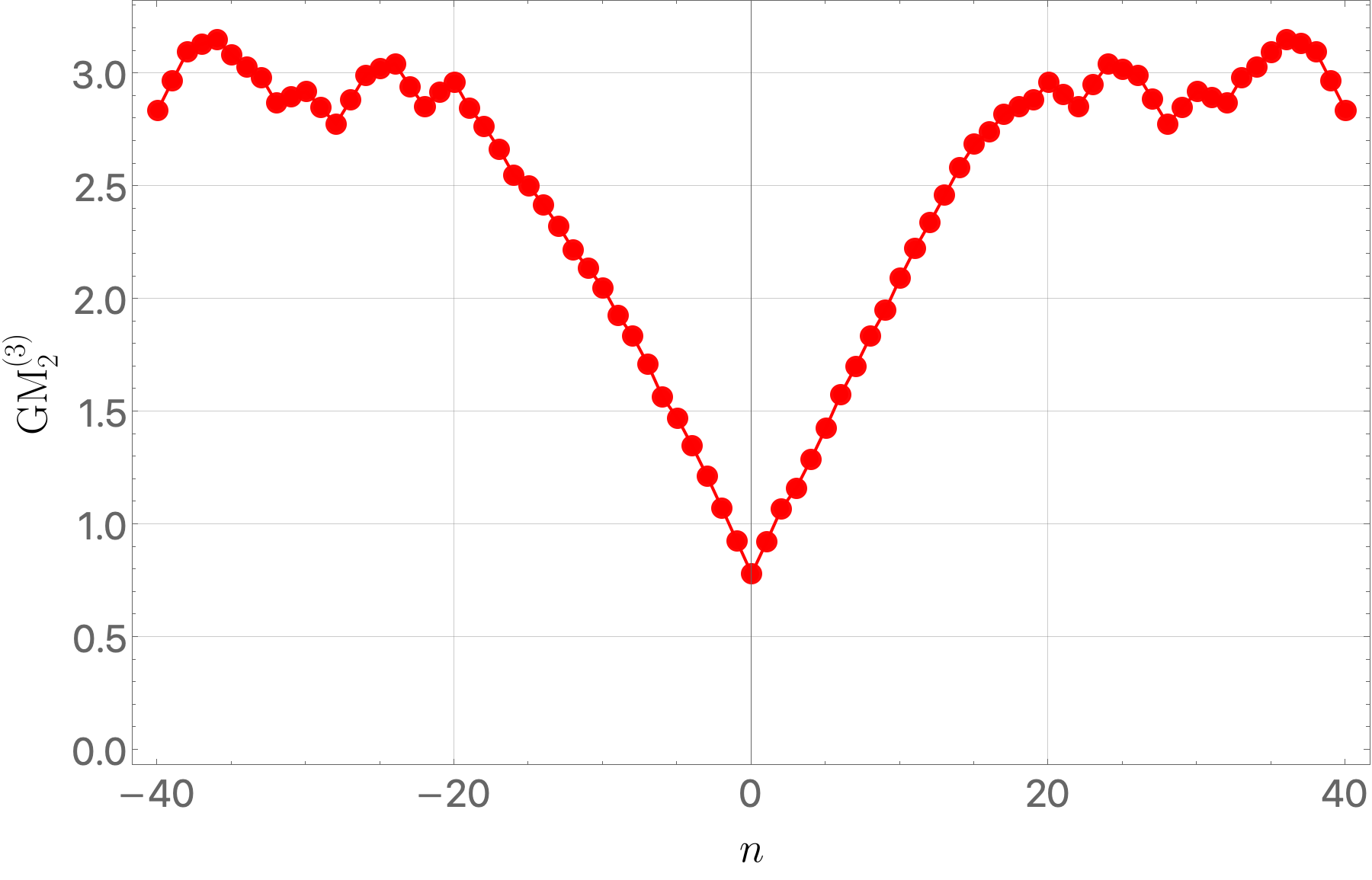}
		\caption{
			\textbf{Genuine multi-entropy $\GM^{(3)}_2$ versus excitation level.} 
			The R\'enyi-2 genuine multi-entropy $\GM^{(3)}_2$ as a function of the excitation level $n$ of (\ref{eq:n-th-state}) for the tripartition shown in Fig.~\ref{fig:setup}(a). The parameters are set to $m=1/10$ and $L=40$. 
			\textbf{Top:} A macroscopic view over a wide range of $n$. Instead of computing all $L^2$ possible values, the data points are appropriately sampled. 
			\textbf{Bottom:} A zoomed-in view around the half-filled ground state ($n=0$), where the data points are densely plotted.
		}
		\label{fig:GM_vs_n}
	\end{figure}
	
	\begin{figure}[h]
		\centering
		\includegraphics[width=0.69\columnwidth]{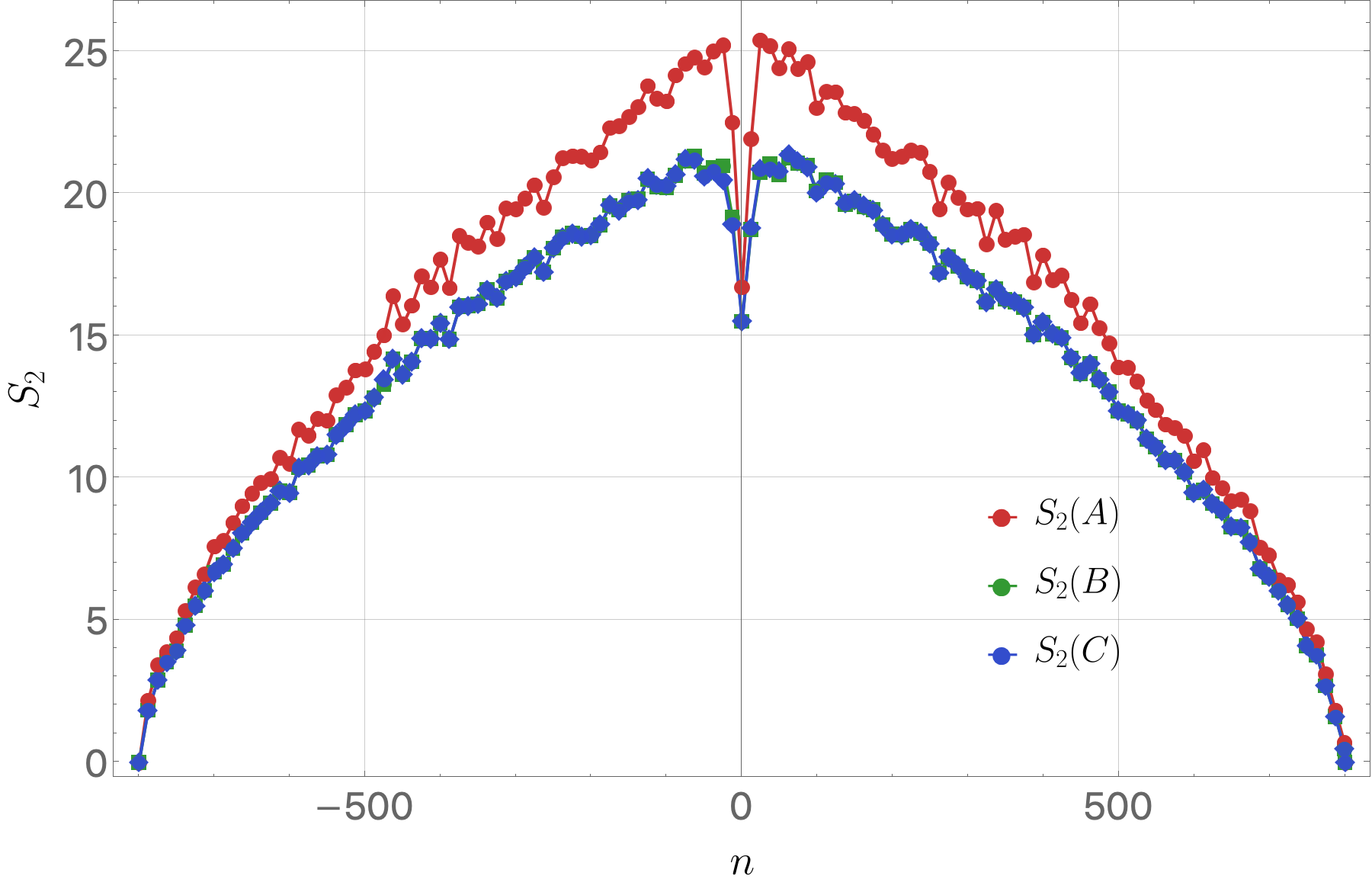} \\
		\vspace{2mm}
		\includegraphics[width=0.69\columnwidth]{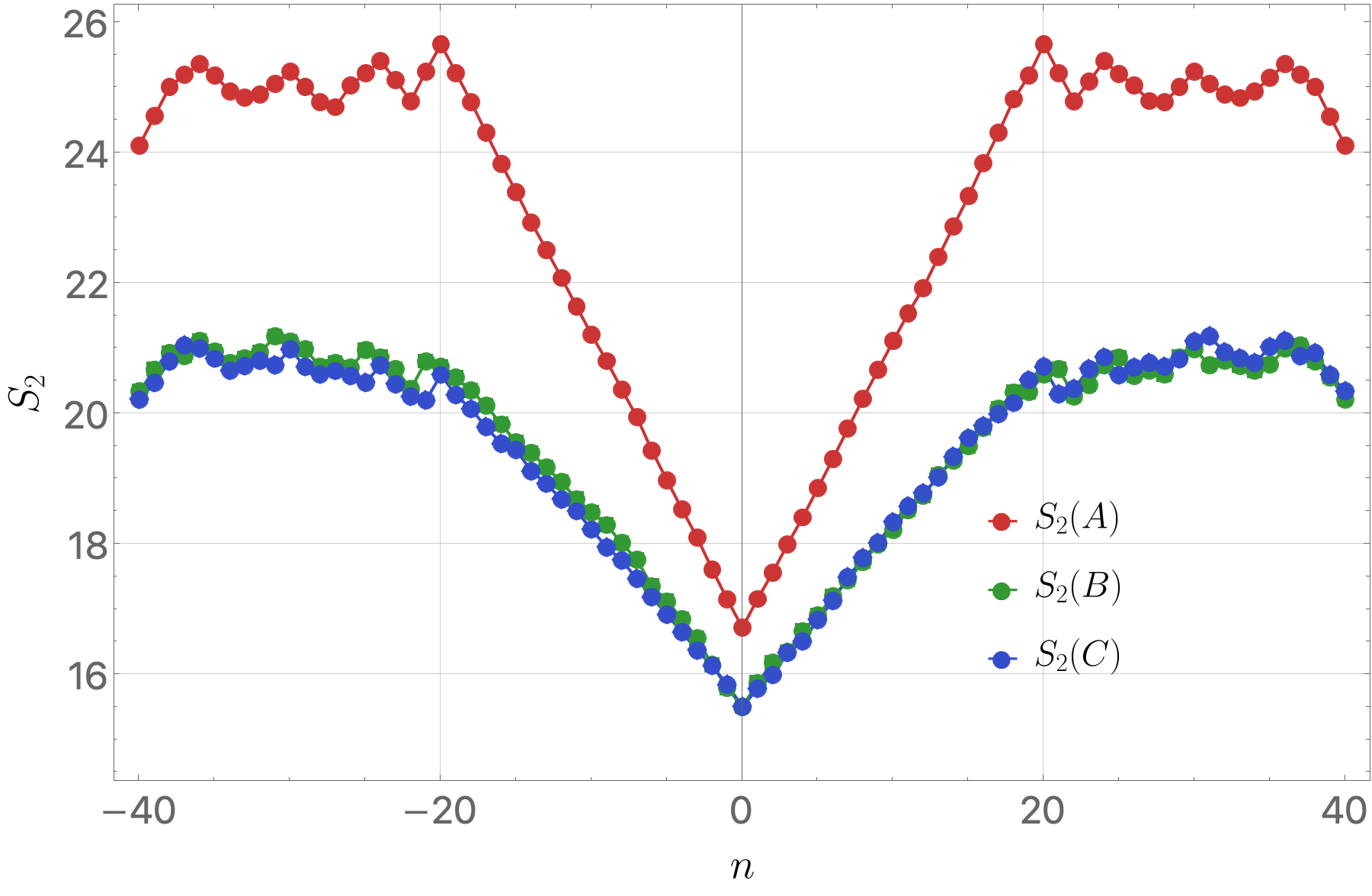}
		\caption{
			\textbf{Entanglement entropy $S_2$ versus excitation level.} 
			The R\'enyi-2 entanglement entropy $S_2$ of subsystems $A$, $B$, and $C$ as a function of the excitation level $n$, using the same parameters and tripartition as in Fig.~\ref{fig:GM_vs_n}. 
			\textbf{Top:} A macroscopic view over a wide range of $n$ with appropriately sampled data points. 
			\textbf{Bottom:} A zoomed-in view around $n=0$, where the data points are densely plotted.
		}
		\label{fig:S2_vs_n}
	\end{figure}

	To provide a brief overview of the entanglement structure for this one-parameter family of states $\{\ket{\Psi_{n}}\}$, in Figures~\ref{fig:GM_vs_n} and \ref{fig:S2_vs_n}, we plot the R\'enyi-2 genuine multi-entropy $\GM^{(3)}_2$ and the R\'enyi-2 entanglement entropies $S_2$ as functions of the excitation level $n$. These quantities are evaluated using the techniques detailed in Appendices~C and E.

%%%%%%%%%%%%%%%%%%%%%%%%%%%%%%%%%%%%%%%%%%%%%%%%%%%%%%%%%%%%%%%%%%%%
\section{Appendix G: Supplementary numerical data and full-scale plots}
%%%%%%%%%%%%%%%%%%%%%%%%%%%%%%%%%%%%%%%%%%%%%%%%%%%%%%%%%%%%%%%%%%%%

This appendix collects supplementary figures that complement the main text.
Some plots in the main text use vertical clipping to highlight the scaling collapse.
Here we provide (i) the corresponding full-scale plots without vertical clipping, and
(ii) representative single-series plots that make rare outliers visible.
These outliers are infrequent and do not affect the typical scaling collapse shown in the main text,
but their microscopic origin remains an interesting open question.

\FloatBarrier
\begin{figure}[!t]
  \centering
  \includegraphics[width=0.75\linewidth]{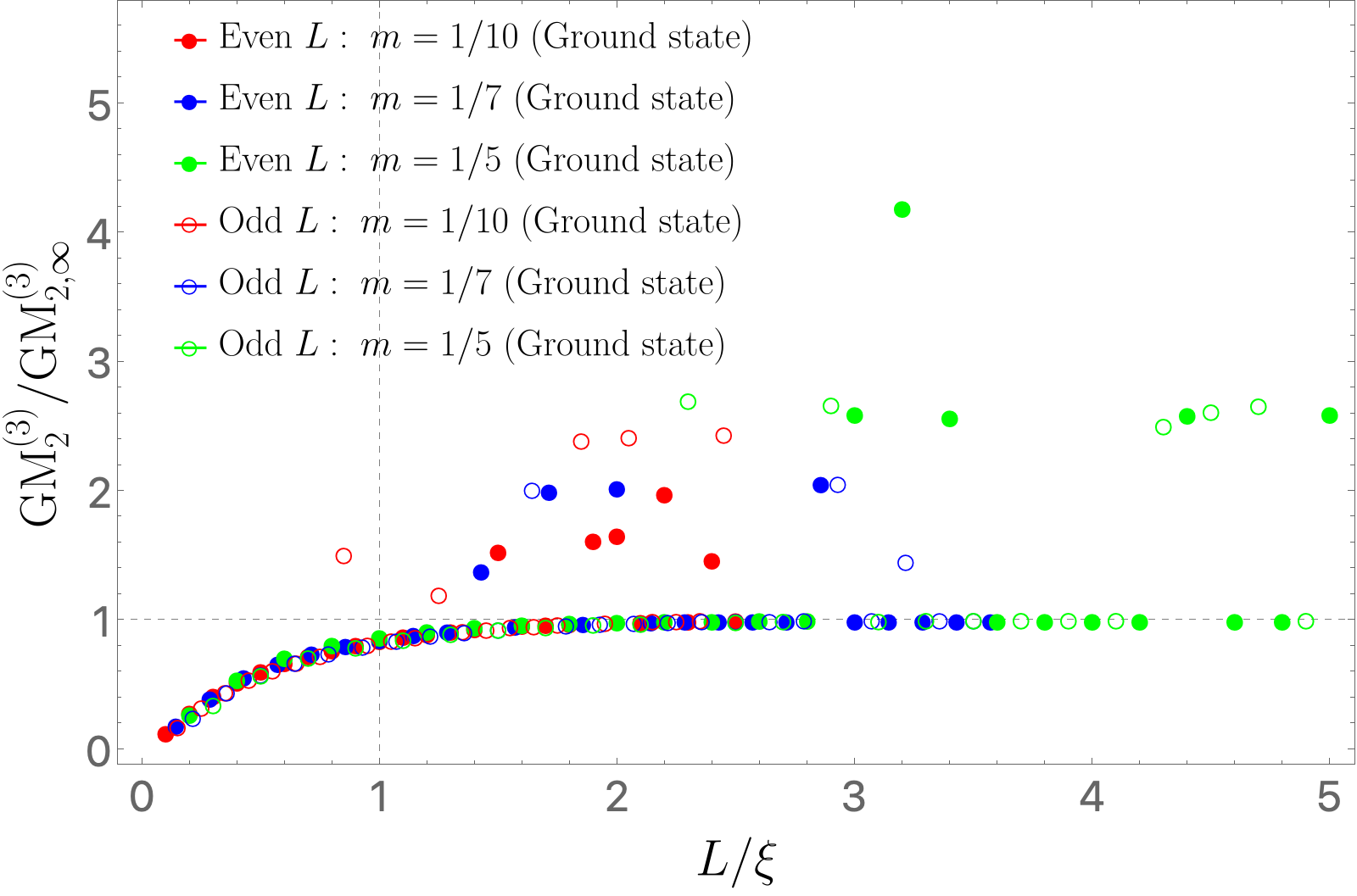}

  \vspace{2mm}

  \includegraphics[width=0.75\linewidth]{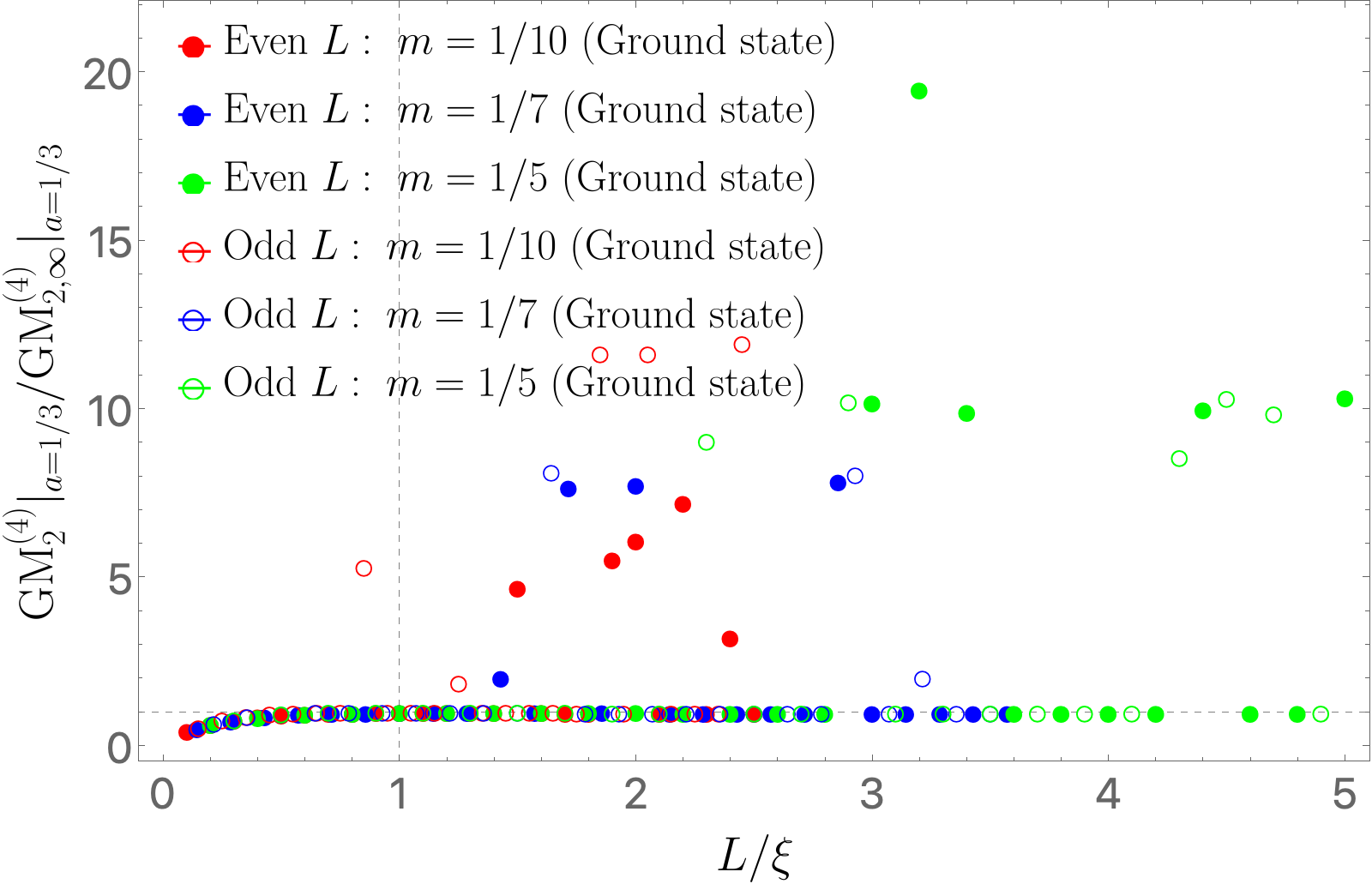}

  \vspace{2mm}

  \includegraphics[width=0.75\linewidth]{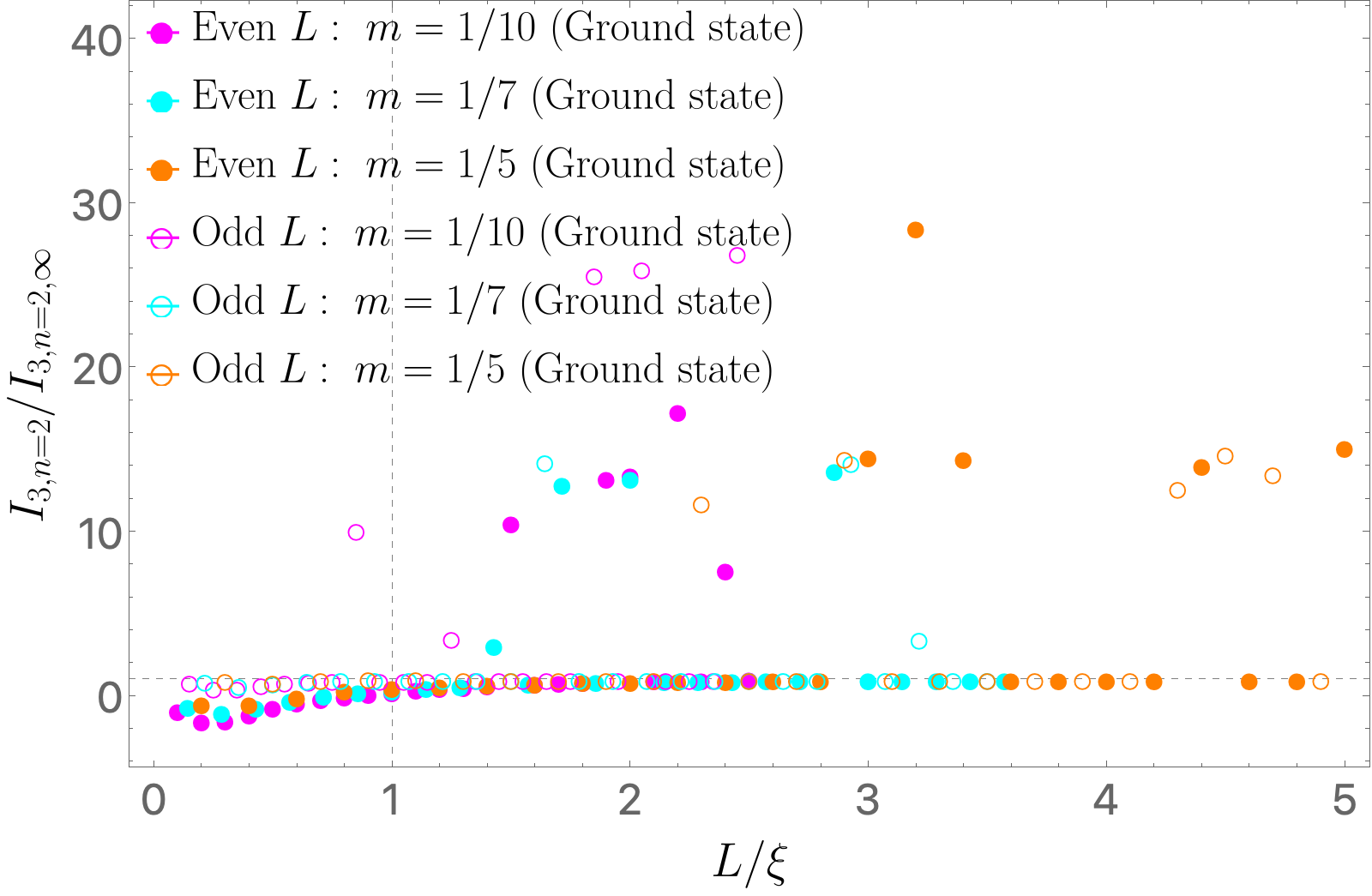}
  \caption{
    \textbf{Junction present: full-scale plots for $\q=3,4$.}
    Full-scale versions of Fig.~\ref{fig:junction_q3}(a) and Fig.~\ref{fig:junction_q4}
    (shown without vertical clipping).
  }
  \label{fig:Fullscale-junction_q34}
\end{figure}

\begin{figure}[!t]
  \centering
  \includegraphics[width=0.78\linewidth]{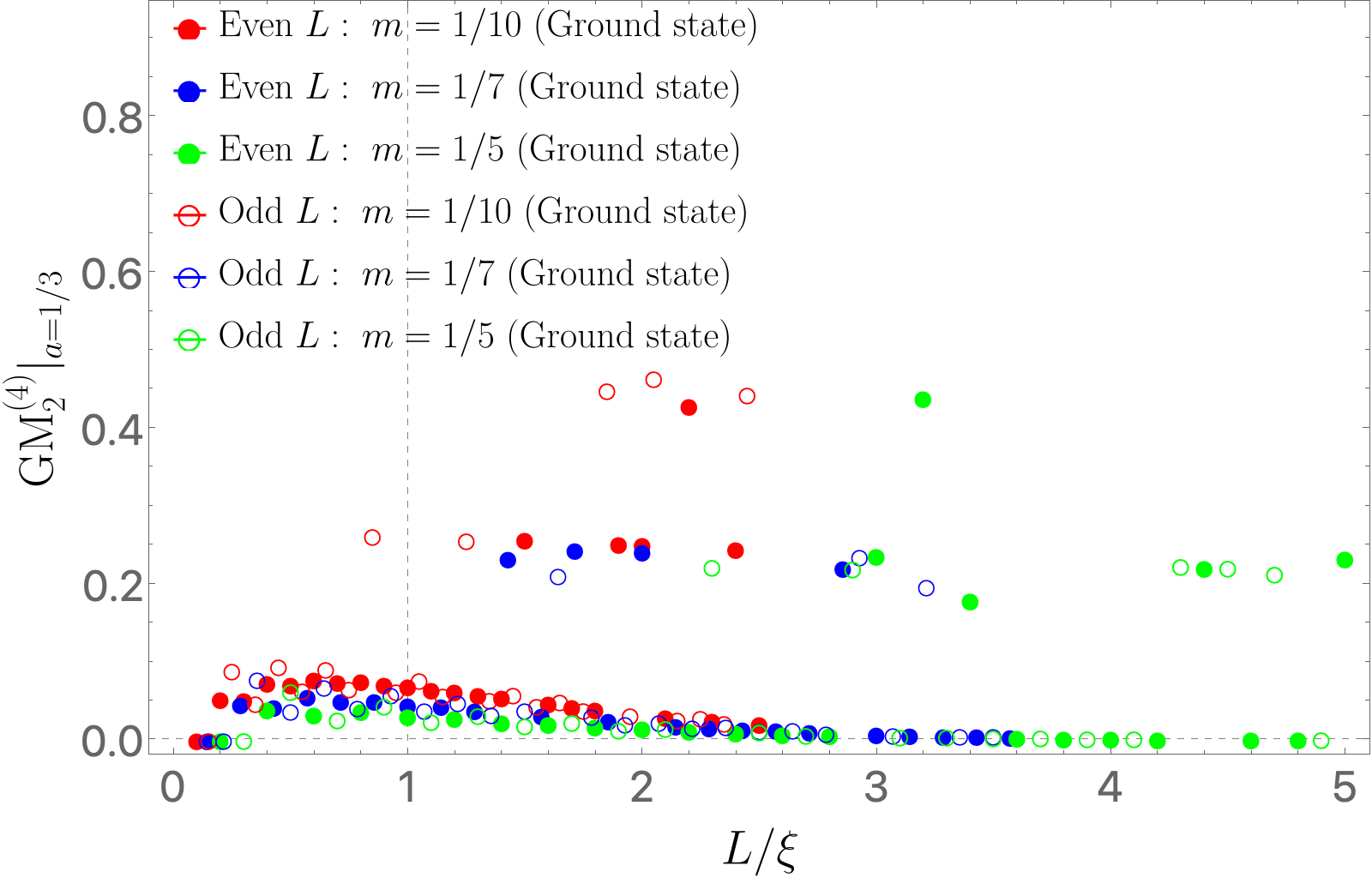}

  \includegraphics[width=0.78\linewidth]{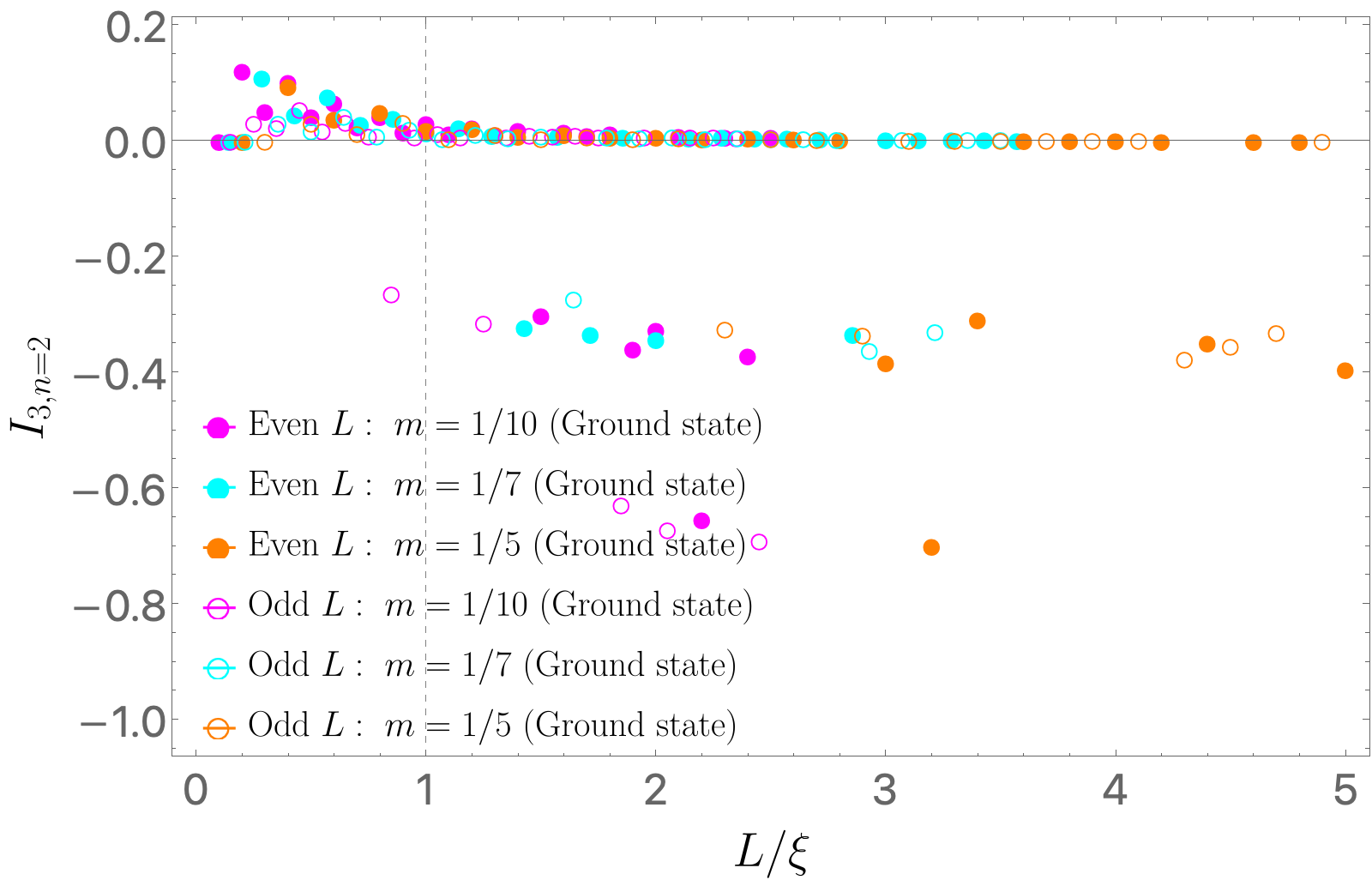}
  \caption{
    \textbf{No junction: full-scale plot.}
    Full-scale version of Fig.~\ref{fig:nojunction} (shown without vertical clipping).
  }
  \label{fig:Fullscale-nojunction}
\end{figure}

\begin{figure}[!t]
  \centering
  \includegraphics[width=0.78\linewidth]{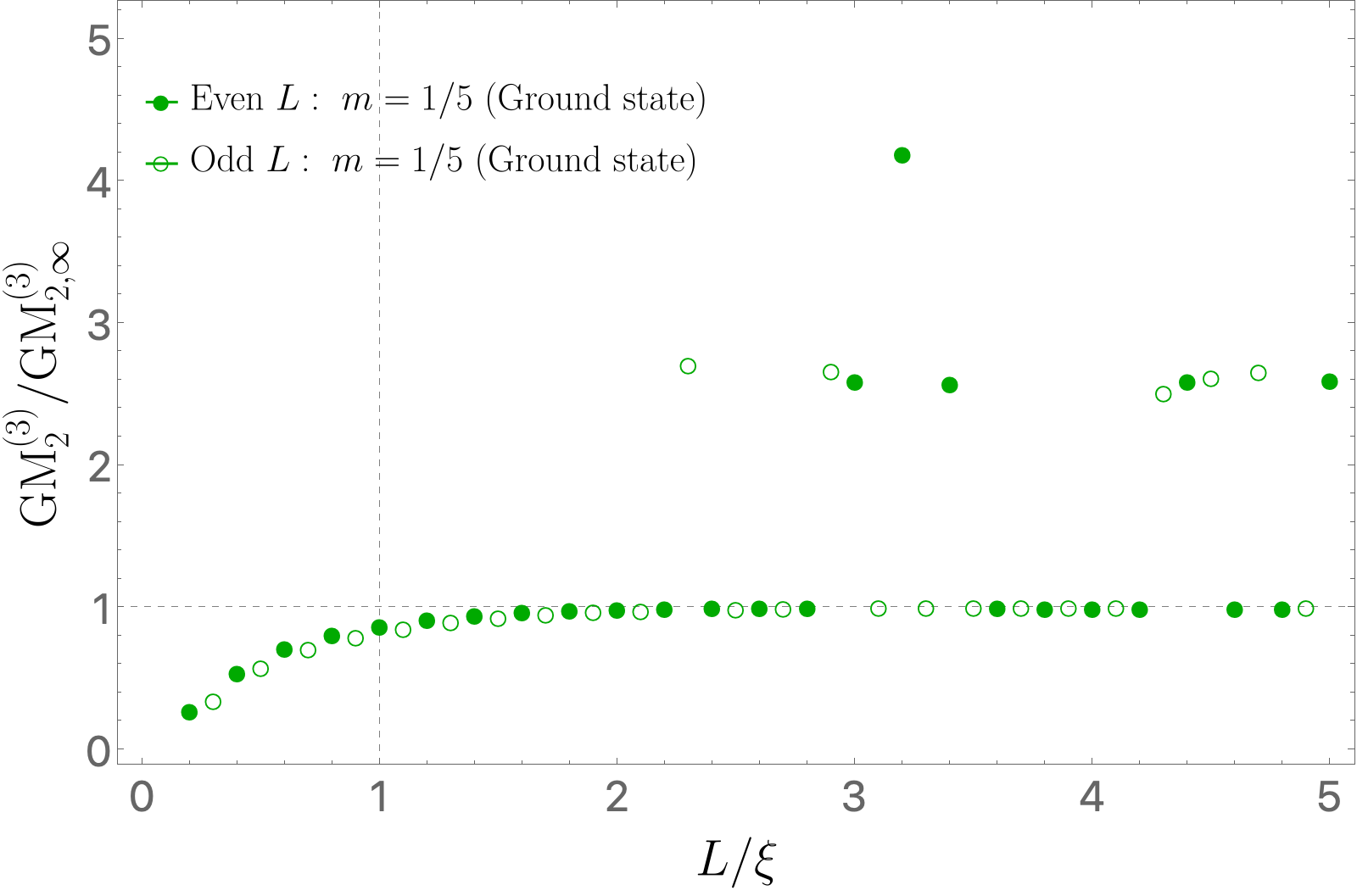}

  \vspace{2mm}

  \includegraphics[width=0.78\linewidth]{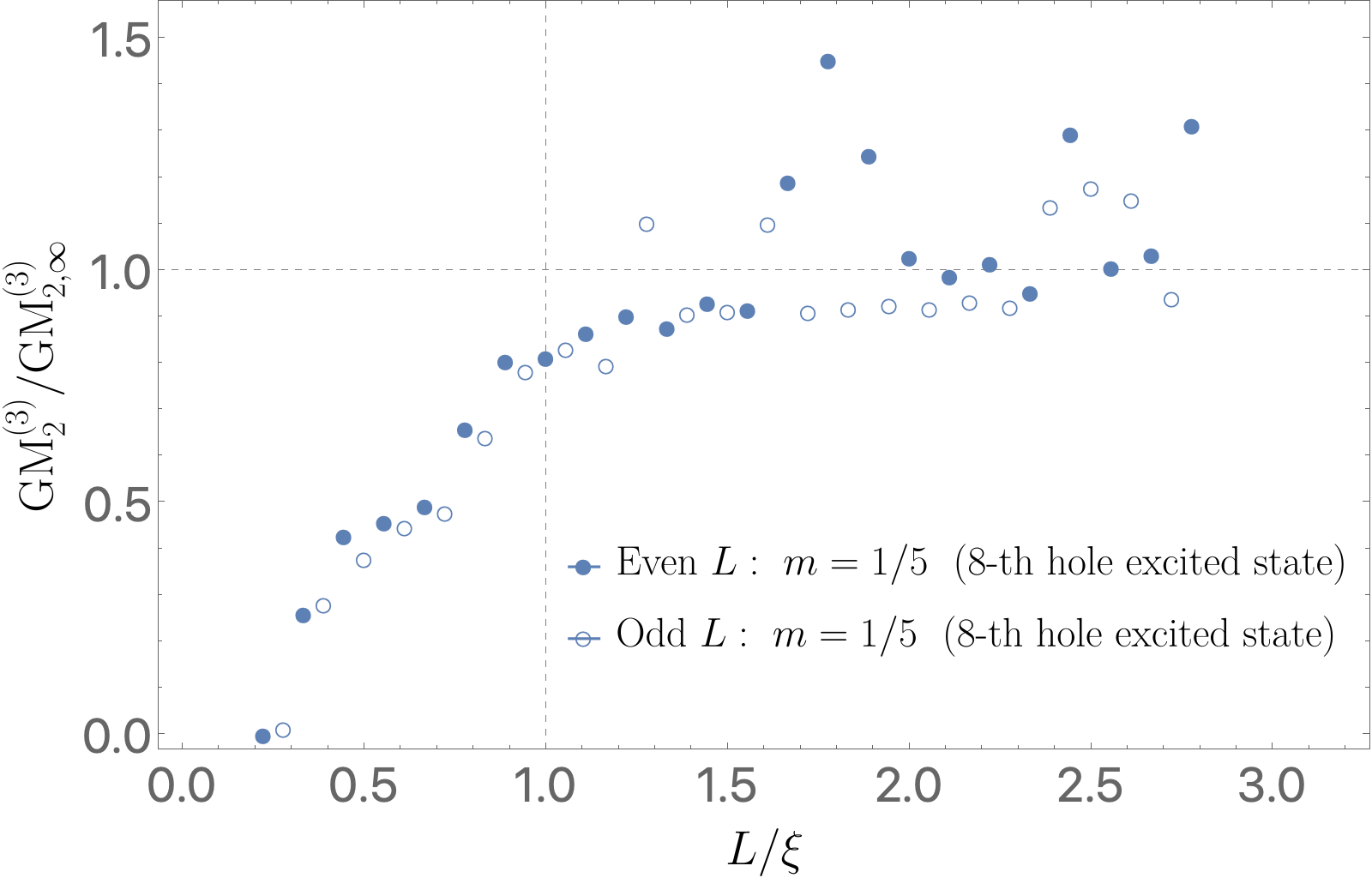}
 \caption{
\textbf{Tripartite junction ($\q=3$): rare outliers.}
Representative single series ($m=1/5$; cf.\ Fig.~\ref{fig:junction_q3}) shown without vertical clipping.
\textbf{Top:} ground state. \textbf{Bottom:} 8th hole excitation.
}
  \label{fig:app_junction_q3}
\end{figure}

\FloatBarrier
\begin{figure}[!t]
  \centering
  \includegraphics[width=0.78\linewidth]{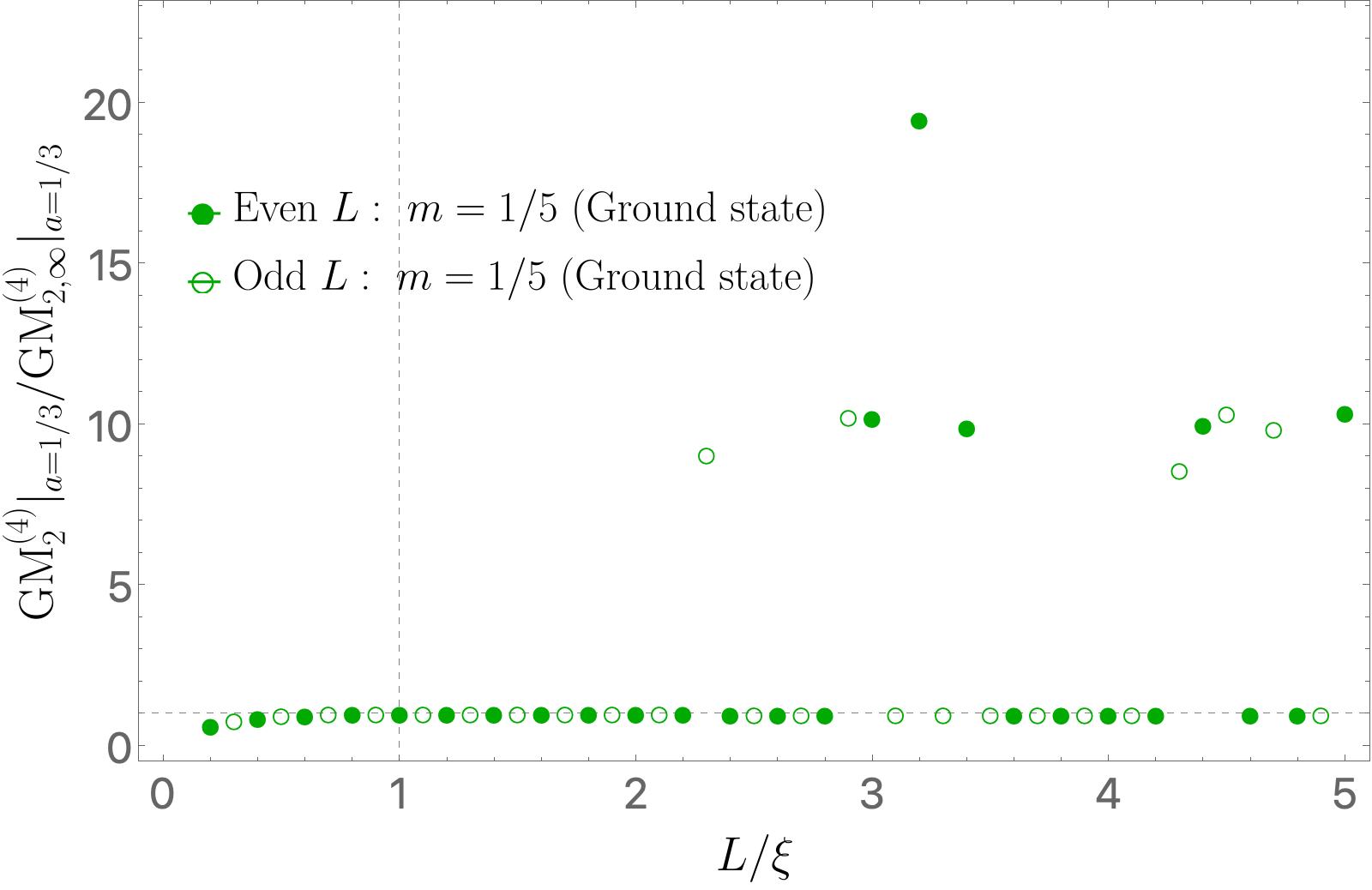}

  \vspace{2mm}

  \includegraphics[width=0.78\linewidth]{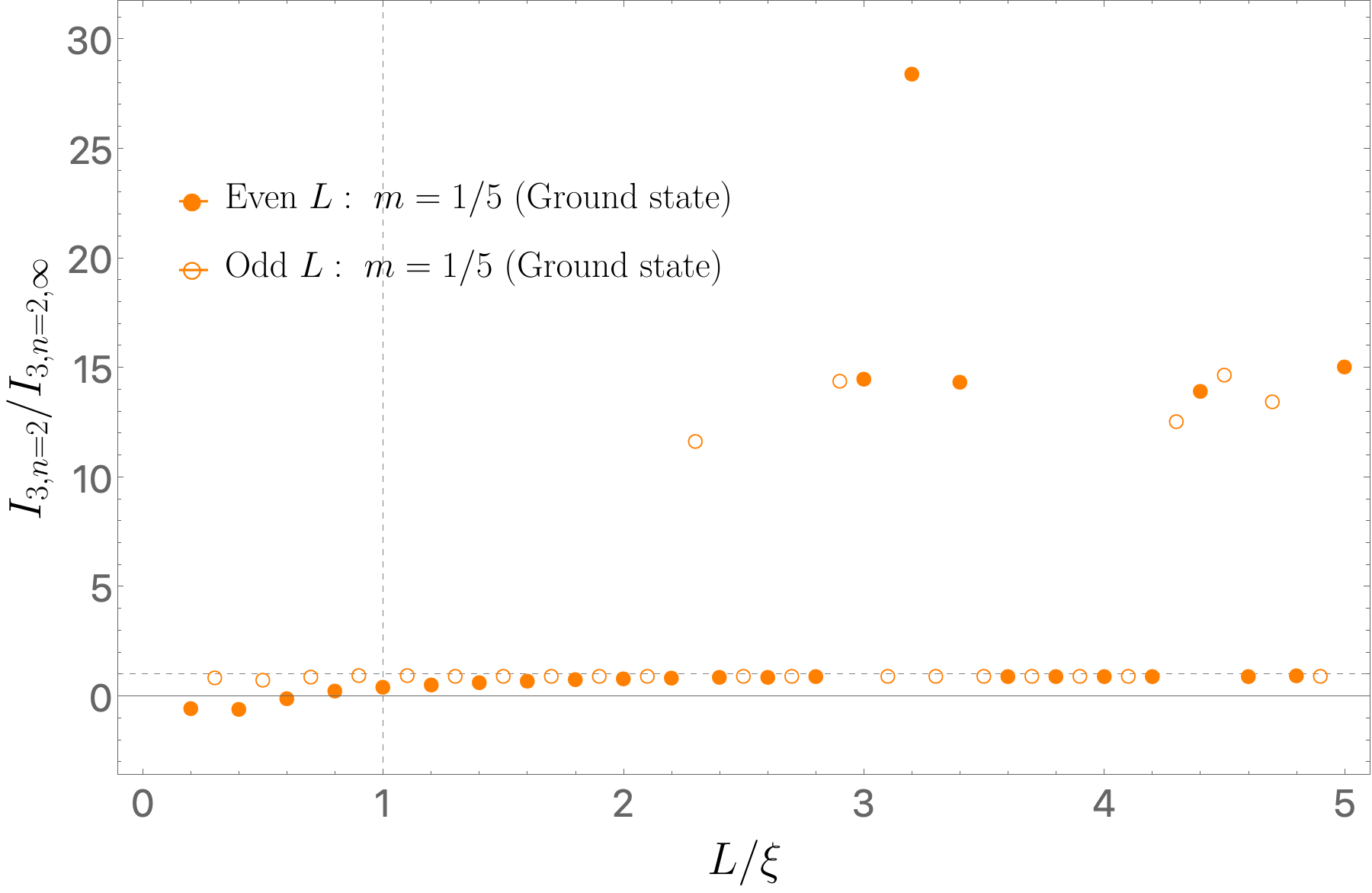}
  \caption{
    \textbf{Quadripartite junction case ($\q=4$): rare outliers in a single series.}
    A representative data series at $m=1/5$ corresponding to Fig.~\ref{fig:junction_q4}
    (shown without vertical clipping).
    \textbf{Top:} $\GM^{(4)}_2|_{a=1/3}$.
    \textbf{Bottom:} $I_{3,n=2}$.
  }
  \label{fig:app_junction_q4}
\end{figure}

\begin{figure}[!b]
% \begin{figure}[!t]
  \centering
  \includegraphics[width=0.78\linewidth]{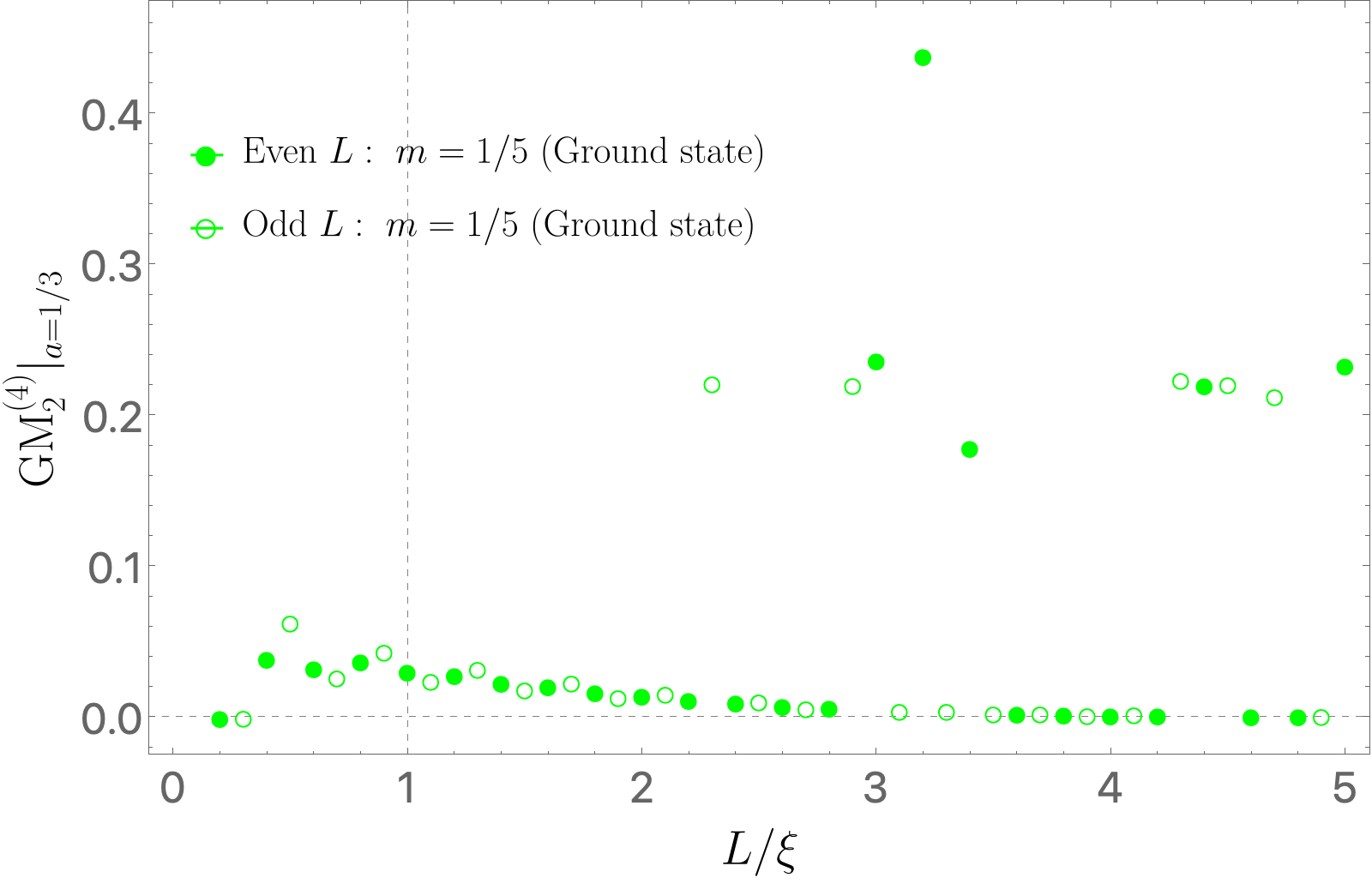}

%  \vspace{2mm}

  \includegraphics[width=0.78\linewidth]{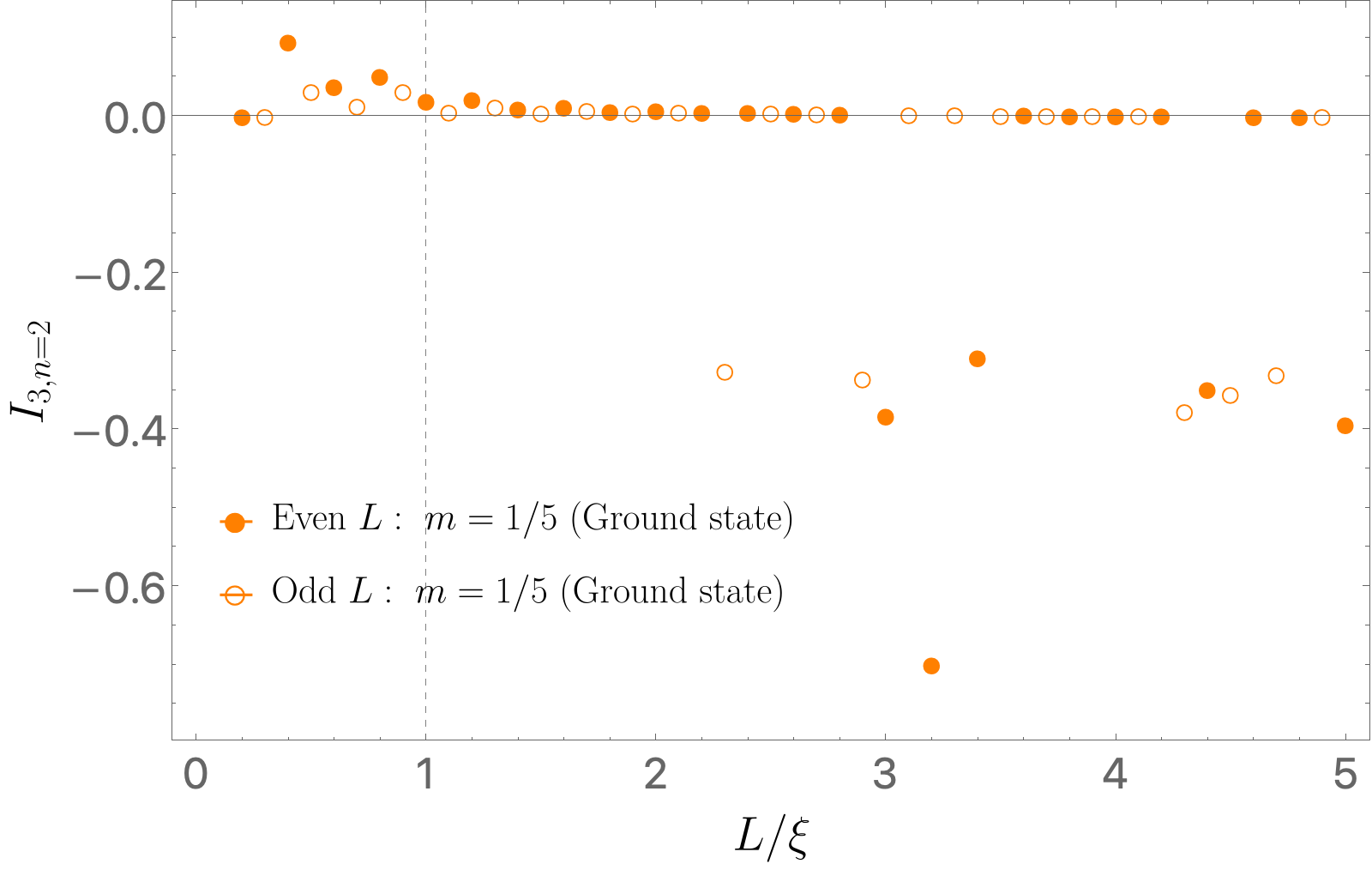}
  \caption{
    \textbf{No-junction case ($\q=4$): rare outliers in a single series.}
    A representative data series at $m=1/5$ corresponding to Fig.~\ref{fig:nojunction}
    (shown without vertical clipping).
    \textbf{Top:} $\GM^{(4)}_2|_{a=1/3}$.
    \textbf{Bottom:} $I_{3,n=2}$.
  }
  \label{fig:app_nojunction}
\end{figure}

\end{document}